\documentclass{aa}
\usepackage{amsmath}
\usepackage[varg]{txfonts}

\def\lsim{\mathrel{\rlap{\lower4pt\hbox{\hskip1pt$\sim$}}
    \raise1pt\hbox{$<$}}}                
\def\gsim{\mathrel{\rlap{\lower4pt\hbox{\hskip1pt$\sim$}}
    \raise1pt\hbox{$>$}}}                

\def\plotonekindaspecial#1{\centering \leavevmode
    \includegraphics[angle=0,width=1.8\columnwidth]{#1}}

\def\plottwo#1#2{\centering \leavevmode
    \includegraphics[angle=0,width=0.98\columnwidth]{#1} \hfil
    \includegraphics[angle=0,width=0.98\columnwidth]{#2}}

\def\plottwoalmostspecial#1#2{\centering \leavevmode
    \includegraphics[angle=0,width=0.9\columnwidth]{#1} \hfil
    \includegraphics[angle=0,width=0.9\columnwidth]{#2}}

\def\plottwosuperspecial#1#2{\centering \leavevmode
    \includegraphics[angle=0,height=0.85\columnwidth]{#1} \hfil
    \includegraphics[angle=0,width=0.97\columnwidth]{#2}}

\def\plottwoveryspecial#1#2{\centering \leavevmode
    \includegraphics[angle=0,width=0.40\columnwidth]{#1} \hfil
    \includegraphics[angle=0,width=0.59\columnwidth]{#2}}

\def\plottwonotsospecial#1#2{\centering \leavevmode
    \includegraphics[angle=0,width=0.98\columnwidth]{#1} \hfil
    \includegraphics[angle=0,width=0.97\columnwidth]{#2}}

\def\plotone#1{\centering \leavevmode
    \includegraphics[angle=0,width=1.0\columnwidth]{#1}}

\def\eps@scaling{1.0}%
\newcommand\epsscale[1]{\def\eps@scaling{#1}}%

\usepackage{marginnote}
\usepackage{color}
\begin{document}

\title{VIMOS Ultra-Deep Survey (VUDS)\thanks{Based on data obtained with the European Southern Observatory Very Large
Telescope, Paranal, Chile, under Large Program 185.A-0791.} : Witnessing the assembly of a massive cluster at $z\sim3.3$\thanks{Table 2 
is only available at the CDS via anonymous ftp to http://cdsarc.u-strasbg.fr (ftp://130.79.128.5) or http://cdsarc.u-strasbg.fr/viz-bin/qcat?J/A+A/}} 
\author{B.~C.\ Lemaux\inst{1} \and O. Cucciati\inst{6,2} \and L. A. M. Tasca\inst{1} \and O. Le F\`evre\inst{1} \and G. Zamorani \inst{2} \and P. Cassata\inst{1,3} \and B. Garilli\inst{4} \and 
V. Le Brun\inst{1} \and D. Maccagni\inst{4} \and L. Pentericci\inst{5} \and R. Thomas\inst{1} \and E. Vanzella\inst{2} \and E. Zucca\inst{2} \and R. Amor\'in\inst{5}
\and S. Bardelli\inst{2}
\and P. Capak\inst{13}
\and L.~P. Cassar\`a\inst{4}
\and M. Castellano\inst{5}
\and A. Cimatti\inst{6}
\and J.~G. Cuby\inst{1}
\and S. de la Torre\inst{1}
\and A. Durkalec\inst{1}
\and A. Fontana\inst{5}
\and M. Giavalisco\inst{14}
\and A. Grazian\inst{5}
\and N.~P. Hathi\inst{1}
\and O. Ilbert\inst{1}
\and C. Moreau\inst{1}
\and S. Paltani\inst{10}
\and B. Ribeiro\inst{1}
\and M. Salvato\inst{15}
\and D. Schaerer\inst{11,9}
\and M. Scodeggio\inst{4}
\and V. Sommariva\inst{6,5}
\and M. Talia\inst{6}
\and Y. Taniguchi\inst{16}
\and L. Tresse\inst{1}
\and D. Vergani\inst{7,2}
\and P.~W. Wang\inst{1}
\and S. Charlot\inst{8}
\and T. Contini\inst{9}
\and S. Fotopoulou\inst{10}
\and R.~R. Gal\inst{18}
\and D.~D. Kocevski\inst{19}
\and C. L\'opez-Sanjuan\inst{12}
\and L.~M. Lubin\inst{17}
\and Y. Mellier\inst{8}
\and T. Sadibekova\inst{20}
\and N. Scoville\inst{13}
}

\institute{Aix Marseille Universit\'e, CNRS, LAM (Laboratoire d'Astrophysique de Marseille) UMR 7326, 13388, Marseille, France \email{brian.lemaux@lam.fr}
\and
INAF--Osservatorio Astronomico di Bologna, via Ranzani,1, I-40127, Bologna, Italy
\and
Instituto de Fisica y Astronom\'ia, Facultad de Ciencias, Universidad de Valpara\'iso, Gran Breta$\rm{\tilde{n}}$a 1111, Playa Ancha, Valpara\'iso Chile
\and
INAF--IASF, via Bassini 15, I-20133,  Milano, Italy
\and
INAF--Osservatorio Astronomico di Roma, via di Frascati 33, I-00040,  Monte Porzio Catone, Italy
\and
University of Bologna, Department of Physics and Astronomy (DIFA), V.le Berti Pichat, 6/2 - 40127, Bologna, Italy
\and
INAF--IASF Bologna, via Gobetti 101, I--40129,  Bologna, Italy
\and
Institut d'Astrophysique de Paris, UMR7095 CNRS,
Universit\'e Pierre et Marie Curie, 98 bis Boulevard Arago, 75014
Paris, France
\and
Institut de Recherche en Astrophysique et Plan\'etologie - IRAP, CNRS, Universit\'e de Toulouse, UPS-OMP, 14, avenue E. Belin, F31400
Toulouse, France
\and
Department of Astronomy, University of Geneva
ch. d'\'Ecogia 16, CH-1290 Versoix, Switzerland
\and
Geneva Observatory, University of Geneva, ch. des Maillettes 51, CH-1290 Versoix, Switzerland
\and
Centro de Estudios de F\'isica del Cosmos de Arag\'on, Teruel, Spain
\and
Department of Astronomy, California Institute of Technology, 1200 E. California Blvd., MC 249-17, Pasadena, CA 91125, USA
\and
Astronomy Department, University of Massachusetts, Amherst, MA 01003, USA
\and
Max-Planck-Institut f\"ur Extraterrestrische Physik, Postfach 1312, D-85741, Garching bei M\"unchen, Germany
\and
Research Center for Space and Cosmic Evolution, Ehime University, Bunkyo-cho 2-5, Matsuyama 790-8577, Japan
\and
Department of Physics, University of California, Davis, 1 Shields Avenue, Davis, CA 95616, USA
\and
University of Hawai'i, Institute for Astronomy, 2680 Woodlawn Drive, Honolulu, HI 96822, USA
\and 
Department of Physics and Astronomy, University of Kentucky, Lexington, KY 40506-0055, USA
\and
Laboratoire AIM, CEA/DSM/Irfu/SAp, CEA-Saclay, F-91191 Gif-sur-Yvette Cedex, France
}

\date{Received March 14th, 2014/ Accepted July 29th, 2014} 
\abstract{ Using new spectroscopic observations obtained as part of the VIMOS Ultra-Deep Survey (VUDS), we performed a systematic search for overdense environments in the early universe ($z>2$) 
and report here on the discovery of Cl J0227-0421, a massive protocluster at $z=3.29$. This protocluster is characterized by both the large overdensity
of spectroscopically confirmed members, $\delta_{gal}=10.5\pm2.8$, and a significant overdensity in photometric redshift members. The halo mass of this protocluster is estimated
by a variety of methods to be $\sim3\times10^{14}$ $\mathcal{M}_{\odot}$ at $z\sim3.3$, which, evolved to $z=0$ results in a halo mass rivaling or exceeding that of the Coma cluster.
The properties of 19 spectroscopically confirmed member galaxies are compared with a large sample of VUDS/VVDS galaxies in lower density field environments at similar redshifts. 
We find tentative evidence for an excess of redder, brighter, and more massive galaxies within the confines of the protocluster relative to the field population, which suggests
that we may be observing the beginning of environmentally induced quenching. The properties of these galaxies are investigated, including a discussion of the brightest protocluster galaxy,
which appears to be undergoing vigorous coeval nuclear and starburst activity. The remaining member galaxies appear to have characteristics that are largely similar 
to the field population. Though we find weaker evidence of the suppression of the median star formation rates among and differences in the stacked spectra of member galaxies 
with respect to the field, we defer any conclusions about these trends to future work with the ensemble of protostructures that are found in the full VUDS sample.} 
\titlerunning{VUDS discovery of a high-redshift protocluster}
\authorrunning{B.~C.\ Lemaux et al.}
\maketitle

\section{Introduction}

Large associations of galaxies provide an excellent laboratory for investigating astrophysical phenomena. The most massive of these associations, galaxy clusters and
superclusters (i.e., clusters of clusters), while rare, are useful not only to constrain the dynamics and content of the universe (e.g., Bahcall et al.\ 2003; Reichardt et al.\ 2013), 
but also to study the evolution of galaxies, since the core of galaxy clusters are the regions of the universe where galaxy maturation occurs most rapidly (e.g., Dressler et al.\ 1984; 
Postman et al.\ 2005). This rapid maturation is a result of the large number of transformative mechanisms that a cluster galaxy experiences, 
mechanisms that are less effective or non-existent in regions of typical density in the universe (e.g., Moran et al.\ 2007). 
The number of processes a cluster galaxy is subject to is, however, both a virtue and a complication for 
studying their evolution. While the signs of transformation and evolution are prevalent among galaxies in clusters that have not already depleted their 
galaxies of gas, the large number of physical processes that are effective in overlapping regimes complicates interpretation. Furthermore,
the effectiveness of such mechanisms appears to have a complex relationship with the halo mass of the host cluster and the dynamics of the galaxies that comprise it, the
density and temperature of the intracluster medium (ICM), local galaxy density, mass of the individual galaxies, and cosmic epoch (e.g., Fujita \& Nagashima 1999; Poggianti et al.\ 2010; 
Lemaux et al.\ 2012; Muzzin et al.\ 2012; Dressler et al.\ 2013). The lower mass counterparts to galaxy clusters, galaxy groups, also suffer the same ambiguities. 

As such, despite nearly a century of study into such associations, the role that environment plays in galaxy evolution and the dominant process or processes that serve
to transform cluster or group galaxies is still unclear. In the local universe, the relationship between environment and galaxy evolution has been revolutionized over the 
past decade with the advent of the Sloan Digital Sky Survey (SDSS). Observations from this survey have been used to great effect to study the properties of both 
groups and clusters and their galaxy content (e.g, G{\'o}mez et al.\ 2003; Hansen et al.\ 2009; von der Linden et al.\ 2010) and have led to insight into the nature 
of environmentally driven evolution in the local universe. However, these studies alone provide only a baseline for studies of cluster and group galaxies in the 
higher redshift universe because, in general, the galaxies populating structures in the low-redshift universe have come to the end of their evolution. Initial 
investigations of clusters beyond the local universe found that the fraction of galaxies that displayed a significant gas content, bluer colors, and late-type (i.e., spiral) 
morphologies increased rapidly with decreasing cosmic epoch (Butcher \& Oemler 1984). Yet, thirty years later, the cause or causes of such a trend have not been 
identified definitively. In intermediate-density environments, such as galaxy groups and pairs or small associations of galaxies, significant progress has been made in the past decade to understand 
the relative effect of such processes on galaxy evolution due to the emergence of spectroscopic surveys covering large portions of the sky in the intermediate-redshift universe 
($z\sim1$, e.g., DEEP2, VVDS, zCOSMOS). While such surveys are typically devoid of massive clusters, a testament to their relative scarcity, the large number of 
spectroscopic redshifts, wide field coverage, and quality of both spectroscopic data and associated ancillary data have led to a variety of insights into the 
nature of galaxy evolution in intermediate-density environments (e.g.,  Cooper at al.\ 2006, 2007, 2008; Cucciati et al.\ 2006, 2010a, 2010b, 2012, Tasca et al.\ 2009; 
Peng et al.\ 2010; Presotto et al.\ 2012; George et al.\ 2012; Knobel et al.\ 2013; Kova{\v c} et al.\ 2014).
 
At similar redshifts, systematic spectroscopic studies of clusters and cluster galaxies are somewhat rare. Surveys of clusters extending to several times the virial radius at 
$z\sim$0.5 (e.g., Treu et al.\ 2003; Dressler et al.\ 2004; Poggianti et al.\ 2006; Ma et al.\ 2008, 2010; Oemler et al.\ 2009, 2013) and of massive groups and clusters 
at $z\sim1$ (e.g., Lubin et al.\ 2009; Jeltema et al.\ 2009; Balogh et al.\ 2011; Muzzin et al.\ 2012; Hou et al.\ 2013; Mok et al.\ 2013, 2014) have begun to provide a somewhat coherent 
picture at these redshifts in which galaxy evolution has a complicated dependence on secular (i.e., mass-related) processes, as well as on both the global and the local environment. However, even at 
such redshifts, the effect of residing in the harsh cluster environment for several Gyr is evident among member galaxies, because the fraction of both red and quiescent 
galaxies is observed to be in excess of that of the field at similar redshifts (e.g., Patel et al.\ 2011; Lemaux et al.\ 2012; van der Burg et al.\ 2013). Going to higher redshifts, the effect of the environment 
should be reversed, inducing rather than suppressing star formation as gas-rich galaxies coalesce in the primeval universe. Indeed, tentative evidence for the reversal of the 
correlation between star formation rate (SFR) and galaxy density has already been found at slightly higher redshifts (Tran et al.\ 2010; Santos et al.\ 2014, though 
see also Santos et al.\ 2013; Ziparo et al.\ 2014). 

Observing the reversal of the SFR$-$density relation, as well as contextualizing the massive, red-sequence galaxies (RSGs) observed at $z\sim1$ in cluster and group environments, 
has motivated recent searches for high-redshift ($z\ga1.5$) clusters (e.g., Henry et al.\ 2010; Gobat et al.\ 2011; Papovich et al.\ 2010; Stanford et al.\ 2012; 
Zeimann et al.\ 2012; Newman et al. 2014) or other overdensities (i.e., protoclusters or protostructures) in the early universe (e.g., Steidel et al.\ 2005; Doherty et al.\ 2010; 
Toshikawa et al.\ 2012; Hayashi et al.\ 2012; Koyama et al.\ 2013; Hodge et al.\ 2013). One
of the main difficulties in such searches, beyond the extreme faintness of the bulk of the member populations of such structures, is the failure of search techniques 
that are widely used at lower redshifts. Traditional techniques, such as searching for overdensities of RSGs or the presence of a hot ICM, are predicated on 
the assumption that a sufficiently long time scale has elapsed over which cluster galaxies can be processed. While these techniques can be used to find the most massive and
oldest structures at any given redshift, such searches are biased against exactly the types of structures where the reversal of the SFR$-$density relation should be most 
apparent. One way of circumventing this bias is to search for overdensities of galaxies lying at the same redshift as estimated by broadband photometry (i.e., photometric 
redshifts), which have now largely supplanted searches for high-redshift overdensities of red galaxies. However, the nature of such overdensities cannot be be characterized well
without dedicated spectroscopic followup.

An alternative technique, which is employed especially for searches of the high-redshift universe, is to perform narrow-band imaging or photometric redshift searches 
around massive radio-loud quasars or other types of powerful active galactic nuclei (e.g., Kurk et al.\ 2004; Miley et al.\ 2004; Venemans et al.\ 2004, 2005; 
Zheng et al.\ 2006; Overzier et al.\ 2008; Kuiper et al.\ 2010, 2011, 2012). Such phenomena are typically associated 
with massive galaxies, which are, in turn, typically associated with galaxy overdensities.
While this technique has been successful in observing large numbers of structures or protostructures in the high-redshift universe, it is not at all clear whether such environments
are typical progenitors of lower redshift clusters or are exceptional in some way, which limits their usefulness in contextualizing results at lower redshifts. Additionally, 
narrow-band and spectroscopic searches of Lyman alpha emitter (LAEs) populations in (somewhat) random regions of the sky have revealed protostructures in the very high-redshift 
universe (e.g., Shimasaku et al.\ 2003; Ouchi et al.\ 2005; Lemaux et al.\ 2009; Toshikawa et al.\ 2012). However, such 
surveys cover rather limited portions of the sky and are only effective at observing overdensities of emission line objects, a 
population that, while being readily observed at high redshift because of the \emph{relative} ease of obtaining redshifts of emission line objects, is the 
subdominant population in the early universe (see, e.g., Shapley et al.\ 2003). As such, the 
structures (or protostructures) found by such searches are wildly inhomogeneous (see the recent review in Chiang et al.\ 2013). This inhomogeneity, combined with a
lack of large, comparable samples of galaxies at more moderate (i.e., field) densities at similar redshifts makes interpreting such structures difficult.

Ideally then, one would require a spectroscopic census of galaxy populations residing in both high- and lower-density environments in the high-redshift universe, representative 
in some way of the overall galaxy population at those epochs. With such a census it should be possible to make distinctions between evolution due to environmental processes and 
those driving overall trends observed in galaxy populations as a function of redshift and to properly connect these galaxy populations to their lower redshift descendants.
The recently undertaken VIMOS Ultra-Deep Survey (VUDS; Le F{\`e}vre et al.\ 2014), an enormous 640-hour spectroscopic campaign with the 8.2-m VLT at Cerro Paranal
targeting galaxies over 1 $\Box^{\circ}$ in three fields at $z>2$,
for the first time provides the possibility of undertaking such a search at these redshifts. Like its predecessors at lower redshift, 
the fields targeted in the VUDS survey are random, albeit well-known, patches of the sky. As mentioned earlier, owing to the magnitude limited nature of
field surveys (e.g., AEGIS, Davis et al.\ 2007; Newman et al.\ 2013; VVDS, Le F{\`e}vre et al.\ 2005, 2013; zCOSMOS, Lilly et al.\ 2007, 2009), the scarcity of red galaxies relative 
to bluer galaxies, and the rarity of massive clusters, environmental studies in field surveys, like VUDS, typically suffer the 
problem of limited dynamic ranges in local densities. Indeed, despite extensive spectroscopy from various surveys in the COSMOS (Scoville et al.\ 2007), CFHTLS-D1, E-CDF-S 
(Lehmer et al.\ 2005) fields, the three fields targeted by VUDS, only a few massive spectroscopically confirmed clusters have been found in these fields at $z<1.5$ 
(Gilli et al.\ 2003; Valtchanov et al. 2004; Guzzo et al.\ 2007; Silverman et al.\ 2008).

However, there are several distinct differences between these surveys and VUDS in the way that they relate to a study of the effect of environment on galaxy evolution due to 
the nature of galaxies being probed. LAEs and other star-forming galaxies at high redshift, both of which are selected in VUDS by virtue of a photometric redshift selection,
are known to be highly clumpy populations (e.g., Miyazaki et al.\ 2003; Ouchi et al.\ 2003, 2004, 2005; Lee et al.\ 2006; Bielby et al.\ 2011; Jose et al.\ 2013), making it 
possible to observe a wide dynamic range of local densities. In the high-redshift universe, 
protostructures comprised of such populations are observed (e.g., Steidel et al.\ 1998; Ouchi et al.\ 2005; Capak et al.\ 2011; Tashikawa et al.\ 2012; Chiang et al.\ 2014) 
and found in simulations (e.g., Chiang et al.\ 2013; Zemp 2013; Shattow et al.\ 2013) to be large in transverse extent. This large extent on the sky allows for sampling a larger number of members 
in a single VIMOS pointing than in traditional multi-object spectroscopic surveys of lower-redshift overdense environments. In addition, as a result of a
photometric redshift selection, galaxies that have more distinguishing features in their SED, i.e., both a continuum break at $\sim$4000 \AA\ and the typical continuum break observed
at the Lyman limit and Lyman$\alpha$, will be more likely to be assigned a accurate photometric redshift and are thus more likely to be targeted. Such a sample will be comprised 
of a mix of quiescent, post-starburst, and starburst populations. These populations are instrumental in the investigation the effect of environment on galaxy evolution.
With this in mind, we performed a systematic search for overdensities of galaxies with secure spectroscopic redshifts in all three VUDS fields. The full results
of this search will be published in a future work. In this paper, we focus on the discovery and study of the most significantly detected \emph{spectroscopic} overdensity
in the CFHTLS-D1 field, Cl J0227-0421, a massive forming cluster at $z\sim3.3$. 

The structure of the paper is as follows. \S\ref{obsnred} provides an overview of the spectroscopic and imaging data available in the CFHTLS-D1 field, as well as the derivation
of physical parameters of galaxies in our sample, with particular attention paid to new observations from the VUDS survey. \S\ref{analysis} describes
the search methodology employed and the subsequent discovery of Cl J0227-0421, along with the estimation of its global properties. 
In \S\ref{memprop} we describe the investigation of the 
properties of the spectroscopically confirmed members of Cl J0227-0421 and compare those properties to galaxies in lower-density environments. Finally, \S\ref{conclusions}
presents a summary of our results. Throughout this paper all magnitudes, including those in the IR, are presented in the AB system (Oke \& Gunn 1983; Fukugita et al.\ 1996). 
We adopt a standard concordance $\Lambda$CDM cosmology with $H_{0}$ = 70 km s$^{-1}$, $\Omega_{\Lambda}$ = 0.73, and $\Omega_{M}$ = 0.27. 

\section{Observations}
\label{obsnred}

Over the past decade and a half, the 0226-04 field has been the subject of exhaustive photometric and spectroscopic campaigns. 
First observed in broadband imaging as one of the fields of the VIMOS VLT Deep Survey (Le F\`evre et al. 2004), this field was subsequently adopted as the first of the ``Deep'' 
fields (i.e., D1) of the Canada-France-Hawai'i Telescope Legacy Survey (CFHTLS)\footnote{http://www.cfht.hawaii.edu/Science/CFHTLS/}.
In this section, we first describe the VIMOS Ultra-Deep Survey (VUDS; Le F{\`e}vre et al.\ 2014) data, which have made the discovery of the protostructure 
reported in this paper possible. We then briefly review other spectroscopic redshift surveys of the field, as well as the associated deep imaging data available
in the CFHTLS-D1 field. For a thorough review of all data available in the CFHTLS-D1 field prior to VUDS, see Lemaux et al. (2013) and 
references therein.

\subsection{Spectroscopic data}
\label{spectra}

The primary impetus for the current study comes from the vast spectroscopic data available in the CFHTLS-D1 field, with a particular reliance on recent 
VIsible MultiObject Spectrograph (VIMOS; Le F{\`e}vre et al.\ 2003) spectroscopic observations taken as part of the VIMOS Ultra-Deep Survey (VUDS; Le F{\`e}vre et al.\ 2014). 
We therefore begin here by a brief discussion of the spectroscopic surveys whose data are utilized for this study. 

\subsubsection{The VIMOS Ultra-Deep Survey}
The observations from which a majority of our results are derived were taken from VUDS, a massive 640-hour ($\sim$80 night) 
VIMOS spectroscopic campaign reaching extreme depths ($i^{\prime}\la25$) of three well-known and well-studied regions of the sky, of which, the CFHTLS-D1 field is one. 
The design, goals, and survey strategy of VUDS are described in detail in Le F{\`e}vre et al.\ (2014) and are thus described here only briefly.
The primary goal for the survey is to measure the spectroscopic redshifts of a large sample of galaxies at redshifts $2\la z \la6$. To this end, unlike 
its predecessors that were magnitude limited, the selection of VUDS spectroscopic targets was performed primarily through photometric redshift cuts, 
occasionally supplemented with a variety of magnitude and color$-$color criteria. These selections were used primarily to maximize the number of galaxies with redshifts likely 
in excess of $z\ga2$ (see discussion in Le F{\`e}vre et al.\ 2014). This selection has been used to great effect, as a large fraction of the galaxies spectroscopically
confirmed in VUDS have redshifts $z\ga2$ (though not all interesting VUDS galaxies are at such high redshifts, see Amor\'in et al.\ 2014). As a result, the number of 
spectroscopically confirmed galaxies at these redshifts in the full VUDS sample rivals or 
exceeds the number of spectroscopically confirmed galaxies from \emph{all other surveys combined at redshifts $z\ga2$}. The main novelty of the VUDS observations
is the depth of the spectroscopy and the large wavelength coverage that is afforded by the 50400s integration time per pointing and per grating with the low-resolution blue 
and red gratings on VIMOS (R=230). This combination of wavelength coverage and depth, along with the high redshift of the sample, allows not only for spectroscopic 
confirmation of the Lyman$\alpha$ emitter (LAE) galaxies, galaxies which dominate other high redshift spectroscopic samples, but also for redshift determination 
from Lyman$\alpha$ (hereafter Ly$\alpha$) and interstellar medium (ISM) absorption in those galaxies that exhibit no emission line features. Thus, the VUDS data allow for a selection of a 
\emph{spectroscopic} volume-limited sample of galaxies at redshifts $2\la z \la6$, a sample that probes as faint as $M^{\ast}$+1 at the redshifts of interest for the study 
presented in this paper (see Cassata et al.\ 2014). The flagging code for VUDS is identical to that of the VIMOS VLT Deep Survey (VVDS; see Le F{\`e}vre et al.\ 2013). Although it has not been, 
to date, tested extensively whether the same statistics as derived for the VVDS flags apply to the VUDS data (though see discussion in Le F{\`e}vre et al.\ 2014), 
we adopt here the same reliability thresholds for secure spectroscopic redshifts in VUDS as for the VVDS. Thus, only those VUDS objects that have 
flag = X2, X3, \& X4, where X=0-3\footnote{X=0 is reserved for target galaxies, X=1 for broadline AGN, X=2 for non-targeted objects that fell serendipitously 
on a slit at a spatial location separable from the target, and X=3 for those non-targeted objects that fell serendipitously on a slit at a spatial location
coincident with the target. For more details on the probability of a correct redshift for a given flag, see Le F{\`e}vre et al.\ (2014)}, 
for which the probability of the redshift being correct is in excess of 75\%, are considered reliable 
(hereafter ``secure spectroscopic redshifts''). In total, spectra of 2395 unique objects were obtained on 
the CFHTLS-D1 field as part of VUDS, with 1534 of those resulting in secure spectroscopic redshifts. This represents only 80\% of the final VUDS data on this 
field, because one VUDS VIMOS quadrant, centered at [$\alpha_{J2000}$, $\delta_{J2000}$] = $[$02:24:36.1, -04:44:58] has yet to be reduced at the time of publication.
For further discussion of the survey design, observations, reduction, redshift determination, and the properties of the full VUDS sample see Le F{\`e}vre et al.\ (2014).

\begin{figure}
\plotone{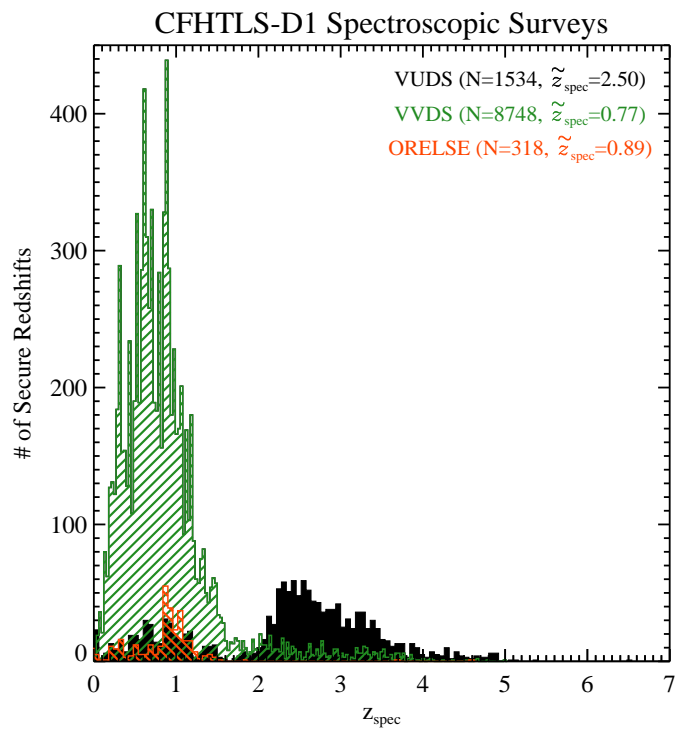}
\caption{Spectroscopic redshift distribution of the 10600 unique objects with secure spectroscopic redshifts (see text) in the CFHTLS-D1 field. The two lower redshift
surveys, VVDS (Le F{\`e}vre et al.\ 2013) and ORELSE (Lubin et al.\ 2009), are shown as hashed green and orange histograms, respectively. The higher 
redshift VUDS (Le F{\`e}vre et al.\ 2014) objects are shown as a black filled histogram. The number of objects with secure spectroscopic redshifts coming from each survey 
along with the median $z_{spec}$ of each sample is shown in the top right corner. For the sake of clarity, the bin size of the histograms for the 
VUDS and ORELSE objects are twice that of the VVDS. Though it is not apparent from the diagram, there is
a tail of galaxies with $z_{spec}>5$ confirmed from the VUDS survey.}
\label{fig:specz}
\end{figure}

\subsubsection{Other spectroscopic data}
The bulk of the lower redshift ($z\la2$) spectroscopy in this field was drawn from observations taken as part of the VVDS ``Deep'' and ``Ultra-Deep'' surveys 
(see Le F{\`e}vre et al.\ 2005, 2013 for details on the survey design and goals) and the Observations of Redshift Evolution in Large Scale Environments 
(ORELSE; Lubin et al.\ 2009) survey. The properties of the spectroscopy available in the CFHTLS-D1 field from these surveys is described
extensively in Lemaux et al.\ (2013) and references therein. These data were used primarily in this study to reject lower redshift 
interlopers and to calibrate and extensively test the SED fitting process that is described in \S\ref{SED}. For the VVDS surveys, the criterion for a 
secure spectroscopic redshift was the same as that of VUDS. For ORELSE, only those objects with quality codes $Q=-1$, 3, \& 4\footnote{$Q=-1$ designates a
stellar spectrum with two or more observed high-quality features, $Q=3$ designates an extragalactic spectrum with at least one high-quality feature observed 
and one feature of marginal quality such that the probability of a correct redshift is $\ge$95\%, and $Q=4$ refers to an extragalactic spectrum with two or more  
high-quality features such that the probability of a correct redshift is $\ge$99\%. See Gal et al. (2008) and Newman et al. (2013) for further details.}
, for which the probability 
that a correct redshift was assigned is in excess of 95\%, were considered secure. Accounting for duplicate observations, a total of 11267, 1120, and 450 spectra were taken
of unique objects in the CFHTLS-D1 field from the VVDS-Deep, VVDS-Ultra-Deep, and ORELSE surveys, respectively, resulting in 7942, 806, and 318 secure spectroscopic
redshifts of unique objects from the three surveys.  Combining all surveys, we have obtained a secure spectroscopic redshift for a total of 10600 unique objects across 
the CFHTLS-D1 field spanning from $0\le z_{spec}\le6.53$. The redshift distributions of those objects with secure spectroscopic redshifts from the three surveys are shown 
in Figure \ref{fig:specz}.

\subsection{Imaging data}
\label{imaging}

Of the plethora of optical imaging data available on the CFHTLS-D1 field, the most relevant for this study is the deep five-band ($u^{\ast}g^{\prime}r^{\prime}i^{\prime}z^{\prime}$)
optical imaging of the entire 1 $\Box^{\circ}$ field observed with Megacam (Boulade et al.\ 2003) as part of the ``Deep'' portion of the CFHTLS survey. 
Model magnitudes (MAG\_AUTO, Kron 1980; Bertin \& Arnouts 1996) were taken from the penultimate release of the CFHTLS data (T0006, Goranova et al.\ 2009) and 
corrected for Galactic extinction and reduction artifacts using the method described in Ilbert et al.\ (2006). The resulting magnitudes reach 5$~\sigma$ point -source completeness limits (i.e., $\sigma_{m}=0.2$) of 26.8/27.4/27.1/26.1/25.7 in the $u^{\ast}g^{\prime}r^{\prime}i^{\prime}z^{\prime}$ bands, respectively, 
sufficient to detect galaxies as faint as $\sim$0.02$L^{\ast}$ at $z=3.3$ (see \S\ref{protoRSG} for the method of estimating $L^{\ast}$).
For further details on the properties of the CFHTLS-D1 imaging and the reduction process, see the CFHTLS TERAPIX
website\footnote{http://terapix.iap.fr/rubrique.php?id\_rubrique=268}, Ilbert et al.\ (2006), and Bielby et al. (2012).

As a compliment to the CFHTLS optical imaging, roughly 75\% of the CFHTLS-D1 field, including the entire area of interest for the present study, was imaged with 
WIRCam (Puget et al.\ 2004) in the near infrared (NIR) $J$, $H$, and $K_s$ bands as part of the WIRCam Deep Survey (WIRDS; Bielby et al.\ 2012). Model magnitudes
were drawn from the T0002 release of WIRDS data\footnote{http://terapix.iap.fr/rubrique.php?id\_rubrique=261} and corrected for Galactic extinction using the method
described in Bielby et al.\ (2012). The resulting magnitudes reach 5$~\sigma$ point source 
completeness limits of 24.7, 24.6, and 24.5 in the $J$, $H$, and $K_{s}$ bands, respectively, sufficient to detect galaxies as faint as $\sim$0.06$L^{\ast}$ at $z=3.3$. 
For further details on the observation, reduction, and characteristics of the WIRDS data see Bielby et al.\ (2012). 

Two generations of imaging with the \emph{Spitzer} Space Telescope were taken on the CFHTLS-D1 field. Initially, a large portion of the CFHTLS-D1 field was imaged at
3.6/4.5/5.8/8.0 $\mu$m from the \emph{Spitzer} InfraRed Array Camera (IRAC; Fazio et al.\ 2004) and at 24$\mu$m from the Multiband Imaging Photometer for 
\emph{Spitzer} (MIPS; Rieke et al.\ 2004) as part of the \emph{Spitzer} Wide-Area InfraRed Extragalactic survey (SWIRE; Lonsdale et al.\ 2003). 
However, these data were too shallow to detect a large majority of the galaxies presented
in this study. Additional \emph{Spitzer}/IRAC data in the two non-cryogenic bands (3.6 \& 4.5 $\mu$m) for the entirety of the field were obtained from the 
Spitzer Extragalactic Representative Volume Survey (SERVS; Mauduit et al.\ 2012). These data, which incorporated the SWIRE data when available, are moderately deeper,
reaching 5$~\sigma$ point- source completeness limits of 23.1 in both $[3.6]$ and $[4.5]$, deep enough to detect a $\sim$0.2$L^{\ast}$ cluster galaxy at $z=3.3$. 
Aperture magnitudes measured within a radius of 1.9$\arcsec$, roughly equivalent to the full-width half-maximum (FWHM) point spread functions (PSFs) of the IRAC 
images in both bands, were drawn from the official SERVS data catalog. These magnitudes were aperture- corrected by dividing the flux density as measured in the aperture
by 0.736 and 0.716 in the 3.6 and 4.5 $\mu$m channels, respectively\footnote{For further details see http://irsa.ipac.caltech.edu/data/ SPITZER/SERVS/docs/SERVS\_DR1\_v1.4.pdf}, 
necessary for matching the model magnitudes of our other optical and NIR (hereafter optical/NIR) imaging. For further details of the reduction of SERVS data for the CFHTLS-D1 field, see Mauduit et al.\ 2012.
The matching of SERVS sources to optical/NIR counterparts from our ground-based imaging was performed by using the known mapping of SWIRE sources (see Arnouts et al.\ 2007) 
when available and nearest-neighbor matching to the combined ground-based optical/NIR catalogs when no SWIRE source was detected at the position of the SERVS source.
In total, 75.5\% of all objects with spectroscopic data were matched to a SERVS counterpart. Even for the highest redshifts probed by the VUDS/VVDS spectroscopy, $z>3$, this
number remains high: a majority (62.5\%) of galaxies with secure spectroscopic redshifts above this limit are matched to a SERVS counterpart.
 
The CFHTLS-D1 field has also been imaged at a variety of other wavelengths with the Very Large Array (VLA), the Giant Millimetre Radio Telescope (GMRT), the
Spectral and Photometric Imaging REceiver (SPIRE; Griffin et al.\ 2010) aboard the \emph{Herschel} Space Observatory (Pilbratt et al.\ 2010), and X-ray Multi-mirror 
Mission space telescope (\emph{XMM-Newton}; Jansen et al.\ 2001). Since these data probe relatively shallowly in the various luminosity functions at $z=3.3$ and will, 
generally, be used only to place upper limits on star formation rates (SFRs), active galactic nuclei (AGN) activity, and intracluster medium (ICM) emission, we refer the reader 
to Lemaux et al.\ (2013) for detailed descriptions of the observations, reduction, and matching of these data.

\subsection{Synthetic model fitting}
\label{SED}

Despite the high density and immense depth of the spectroscopic coverage in the CFHTLS-D1 field, a majority of the objects in the field that are detectable to the 
depth of our imaging data were not targeted with spectroscopy. For these objects, information can only be obtained through fitting to their spectral energy distributions
(SEDs) in the observed-frame optical/NIR broadband photometry. To derive redshifts from the photometric data for untargeted objects, as well as their associated 
physical parameters, e.g., stellar masses, mean luminosity-weighted stellar ages, and SFRs, we utilized the package Le Phare\footnote{http://cfht.hawaii.edu/$\sim$arnouts/LEPHARE/lephare.html}
(Arnouts et al.\ 1999; Ilbert et al.\ 2006, 2009) in a method identical to the one described in Lemaux et al.\ (2013). The process for deriving physical parameters for galaxies
that have been spectroscopically targeted was similar to the one in Lemaux et al.\ (2013) with a few minor modifications that are described in Appendix A. For some analysis,
similar fitting was performed on VUDS rest-frame near-ultraviolet (NUV) spectra using the Galaxy Observed-Simulated SED Interactive Program (GOSSIP; Franzetti
et al.\ 2008). The details of all synthetic model fitting to the photometric and spectroscopic detail, as well as discussions of the effect of various assumptions 
made for these fitting processes, are discussed in detail in Appendix A. In Figure \ref{fig:speczvsphotoz} we show a comparison of the photometric redshifts derived 
from the CFHTLS and WIRDS (hereafter CFHTLS/WIRDS) imaging and their associated spectral redshifts for those galaxies with secure spectroscopic redshifts. Of particular
importance to this work is the true redshift distribution of objects with $z_{phot}>3$, which are almost always (82.7\% of the time) at $z_{spec}>3$. The large
majority of cases where a galaxy is at $z_{spec}>3$ and the photometric redshift estimation failed miserably were when the galaxy was wrongly classified as a star
or the Ly$\alpha$ break was mistaken for the Balmer/4000\AA\ break, which placed the galaxy at very low redshifts. These failures, and similar failures at lower redshift,
produce the parabolic shape seen in the bottom panel extending across the entire redshift range (i.e., these are galaxies which were assigned a $z_{phot}\sim0$).

\begin{figure}
\plotone{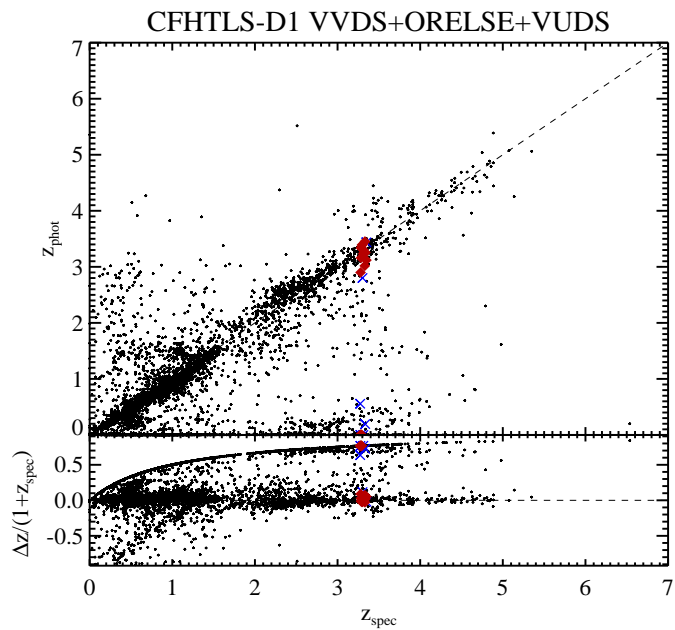}
\caption{Comparison of photometric redshifts as derived from eight-band ground-based optical/NIR imaging and spectroscopic redshifts for those objects with secure 
spectroscopic redshifts (see \S\ref{spectra}). In the bottom panel $\Delta z\equiv (z_{spec}-z_{phot})$. Members of the Cl J0227-0421 protostructure (see \S\ref{protocluster}) 
with secure spectroscopic redshifts are denoted in both panels by red diamonds, those with less secure spectroscopic redshifts are shown as blue Xs. 
For an discussion of the relevance of this comparison for this study and an explanation of the parabolic feature seen in the bottom panel see the text 
at the end of \S\ref{SED}.} 
\label{fig:speczvsphotoz}
\end{figure}

\section{An exploration into the role of environment in VUDS}
\label{analysis}

\begin{figure}
\epsscale{1}
\plotone{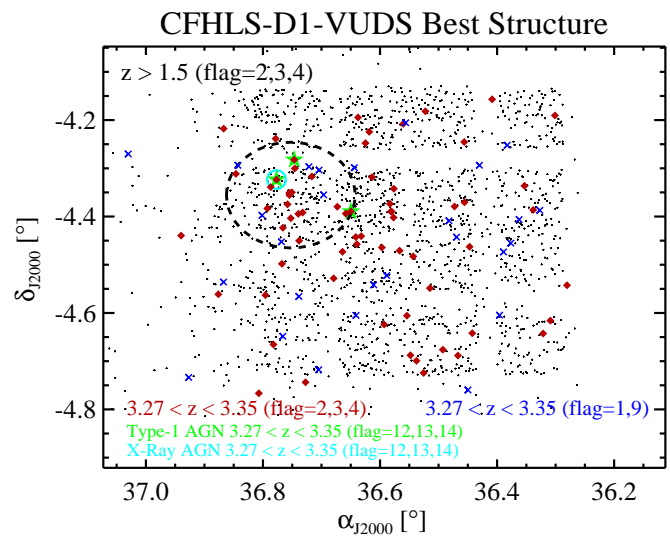}
\caption{Sky distribution of galaxies in the redshift range of the most significantly detected spectroscopic protostructure in the CFHTLS-D1 field (Cl J0227-0421). 
Galaxies with secure spectroscopic redshifts are plotted as red diamonds, those with less secure spectroscopic redshifts are shown as blue Xs. 
Green stars denote those galaxies hosting a type-1 AGN and the cyan circle denotes the lone X-ray AGN host at these redshifts. Plotted in the
background are all galaxies in the CFHTLS-D1 field with secure spectroscopic redshifts $z_{spec}>1.5$. The dashed circle designates 
3 $h^{-1}_{70}$ Mpc from the adopted center of the protostructure and is used in conjunction with the redshift range to define membership (see \S\ref{protocluster}).
The VIMOS footprint is clearly visible, and coverage gaps in the CFHTLS-D1 spectroscopy are apparent throughout the field and to the north of the 
protostructure.} 
\label{fig:LSS}
\end{figure}

We begin the exploration by briefly describing the search technique used to find spectroscopic overdensities of galaxies in the VUDS survey. Though the search technique is broadly similar
in all fields, we limit ourselves here to the search as performed on the CFHTLS-D1 field and defer the discussion of the search in the two other fields for future work (though see Cucciati et al.\ 2014 for a 
discussion of the most significant overdensity in the COSMOS field). The methodology used for the search, along with an involved discussion of the purity and completeness 
of the overdensities found in all VUDS fields, will also be described in a future paper since here we are concerned with only the most significant of the overdensities
in the CFHTLS-D1 field. 

\begin{figure*}
\epsscale{1}
\plottwosuperspecial{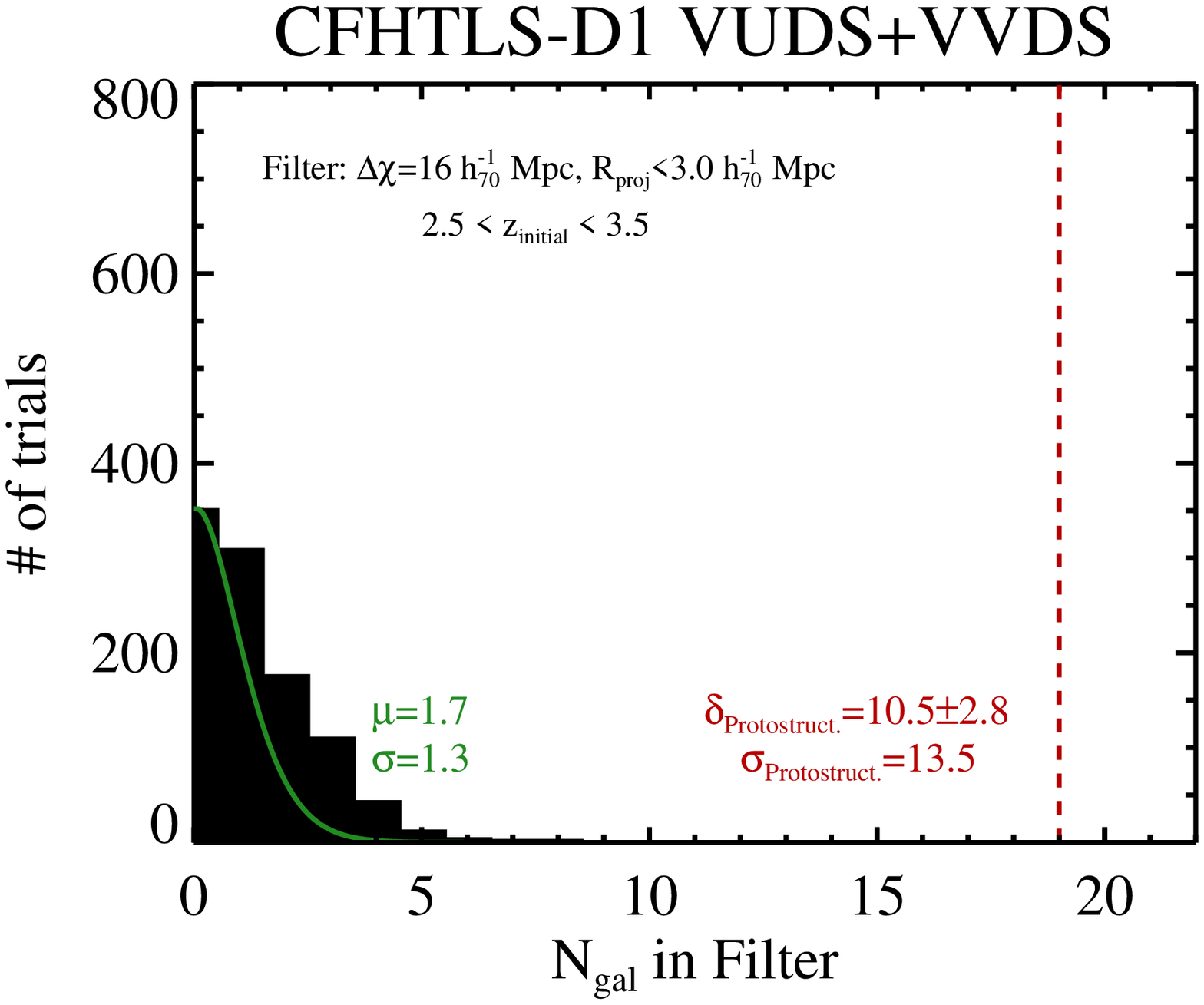}{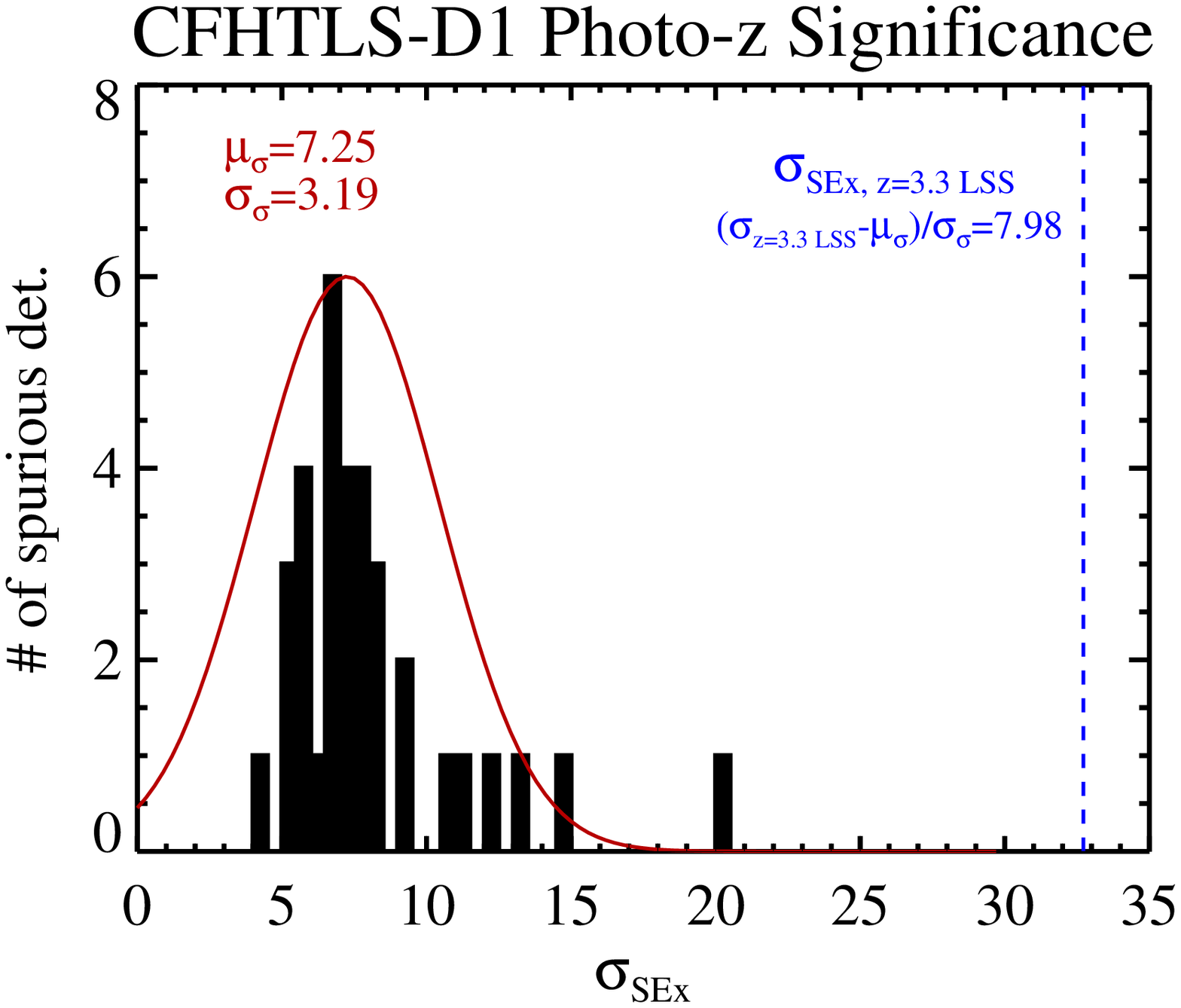}
\caption{\emph{Left:} Spectroscopic overdensity of Cl J0227-0421. Plotted in black is the histogram of the number of galaxies
with secure spectroscopic redshifts falling within a filter of the dimensions listed ($\chi$ refers to proper distance) for 1000 observations 
of random locations and central redshifts ($2.5<z_{spec}<3.5$) across the CFHTLS-D1 field avoiding gaps in spectroscopic coverage. 
The solid green line shows the best-fit Poisson distribution with the numbers to the left denoting the best-fit parameters. The number of members of Cl J0227-0421
with secure spectroscopic redshifts within the filter bounds is plotted as a red vertical dashed line. The Cl J0227-0421 galaxy overdensity, 
$\delta_{gal}\equiv (N_{gal,Proto-Struct.}-\mu)/\mu$, is shown to the left of the vertical dashed line along with the formal significance of the 
spectroscopic overdensity, $\sigma_{Proto-Struct.}\equiv (N_{gal,Proto-Struct.}-\mu)/\sigma$.
\emph{Right:} Photometric redshift galaxy overdensity of Cl J0227-0421. The black histogram shows the Source Extractor (SEx) significance distribution 
of spurious density peaks in the CFHTLS-D1 field (see \S\ref{analysis}). The solid red line shows the best-fit Gaussian distribution to
the significance distribution of the spurious peaks and the associated best-fit parameters are shown above this line. The horizontal
dashed blue line denotes the SEx significance of the photometric redshift galaxy overdensity in Cl J0227-0421, while the 
number directly to the left of the line gives the formal significance of the overdensity that accounts for spurious density peaks.}
\label{fig:sigLSS}
\end{figure*}

The search was performed as follows. All unique galaxies with secure spectroscopic redshifts in the CFHTLS-D1 field 
(see \S\ref{spectra}) were combined into a single catalog, and this catalog was used to generate density maps of secure spectroscopic objects using the methodology of Gutermuth et al. (2005).
To be considered a legitimate overdensity, referred to hereafter by the sufficiently ambiguous term ``protostructure'', we required seven concordant redshifts within a 
circle of radius 2 $h_{70}^{-1}$ proper Mpc at the redshift of the source and a maximum distance between galaxies along the line of sight of 25 $h_{70}^{-1}$ proper Mpc (equivalent to roughly
$\Delta v\sim5000-8500$ km s$^{-1}$ or $\Delta z\sim0.06-0.12$ at the redshifts considered here). This size is well matched to the spatial and redshift extent of both simulated and observed high-redshift
protostructures. We note here that we make no requirement or claim that these protostructures be gravitationally bound, but are instead interested only in their being significantly dense 
relative to the field so as to increase the chance of observing signatures of environmentally-driven evolution. The significance of each spectral overdensity was determined by
``observing'' 1000 protostructure-sized volumes in random locations over the CFHTLS-D1 field (avoiding coverage gaps) at a random central redshift between $2.5<z<3.5$ (see Figure \ref{fig:sigLSS}). 
To date, 13 such protostructures have been discovered in the CFHTLS-D1 field, of which the most significant and the subject of this paper, a protostructure at $z\sim3.3$, is shown in Figure \ref{fig:LSS}. 

\begin{figure*}
\epsscale{1.5}
\plotonekindaspecial{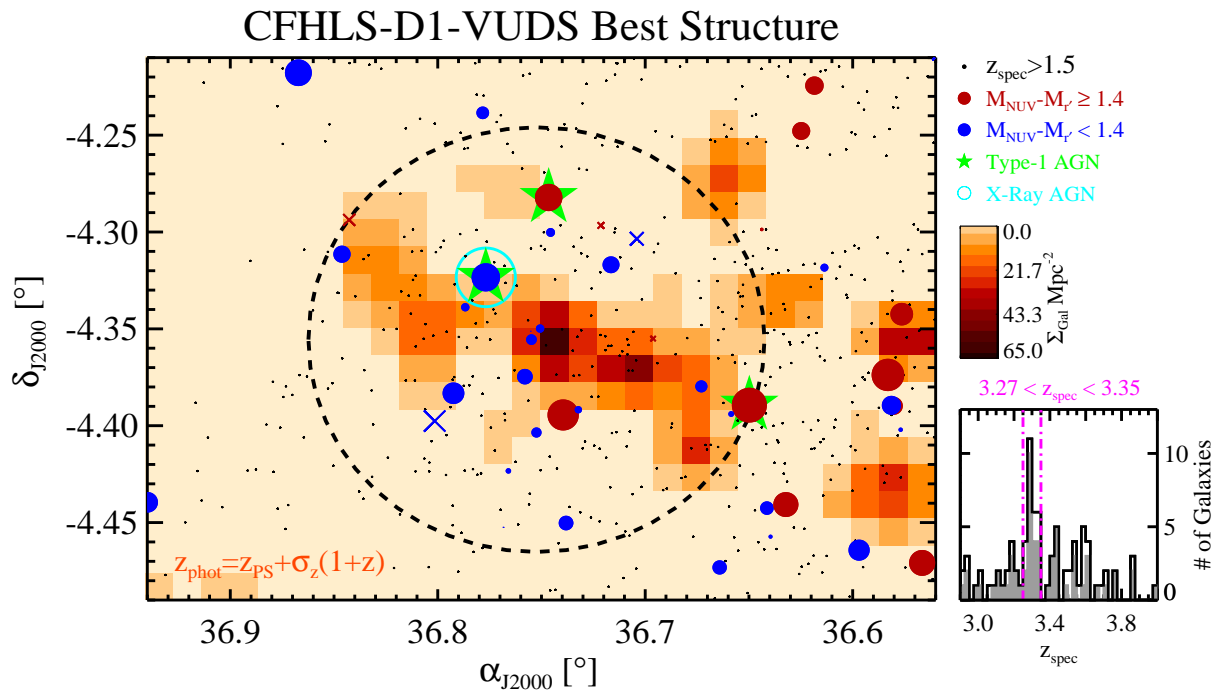}
\caption{\emph{Left:} Zoom-in of the sky distribution of all galaxies in the redshift range of the same protostructure that is shown in Figure \ref{fig:LSS} (Cl J0227-0421). This protostructure 
is also the most significantly detected protostructure in photometric redshift overdensity. The photometric density map, generated using the methodology of \S\ref{analysis}, is shown in the
background, with the scale bar denoting the photometric redshift galaxy density. As in Figure \ref{fig:LSS}, the dashed circle designates
3 $h^{-1}_{70}$ Mpc from the adopted center of the protostructure. Blue and red symbols show galaxies at the redshift of the protostructure differentiated
by their $M_{NUV}-M_{r^{\prime}}$ colors (where the delineation point was set roughly at the color of a 200 Myr old stellar population, see \S\ref{protoRSG}) and are logarithmically scaled ($\log_{4}$) by their stellar mass (see \S\ref{CMDnCSMD}). Filled circles denote
galaxies with secure spectroscopic redshifts, while Xs denote those galaxies with less secure spectroscopic redshifts. Plotted in the background
are all galaxies with $z_{spec}>1.5$. Several massive, redder galaxies are observed in the bounds of the protostructure and extended filamentary structure can be seen to the west-southwest of the central protostructure galaxy concentration.
\emph{Bottom right:} Spectroscopic redshift distribution of galaxies within 3 $h^{-1}_{70}$ Mpc of the
protostructure center. The shaded histogram displays only those galaxies with secure spectroscopic redshifts, while the histogram plotted with a
solid black line also includes those galaxies with less secure spectral redshifts. The redshift bounds defining membership are marked by magenta
dot-dashed lines.}
\label{fig:LSSwdens}
\end{figure*}

Generated \emph{a posteriori} were photometric redshift density maps of all galaxies within $\Delta z_{phot}\pm0.02(1+z_{spec})$ of the 
spectroscopic redshift bounds of each protostructure using the same methodology as used to create the spectral density maps. 
While we did not require an overdensity of photometric redshift sources to consider a grouping of galaxies a protostructure, these 
maps served to lend credence to the overdensity seen in the spectroscopic data and to more fully probe the large scale structure (LSS) of
the galaxy overdensity. The latter is especially important because, as mentioned earlier, both LAEs and other star-forming galaxies 
at high redshift are highly clustered populations, and in a single VIMOS pointing, roughly only 20\% of objects with photometric redshifts at $z_{phot}>2$ can be 
targeted with spectroscopy. Source Extractor (Bertin \& Arnouts 1996) was run on each density map to measure  
significances relative to the background of detections in all density maps. These detections were cross-correlated with the 
spectral density maps to look for spurious density peaks\footnote{More specifically, those photometric redshift source overdensities that did not have a spectroscopic
overdensity sufficient to fulfill the criteria set by the protostructure filter centered anywhere within
1 $h_{70}^{-1}$ Mpc of the peak pixel of the photometric redshift overdensity.}, which was in turn used to define a
significance threshold for photometric redshift galaxy overdensities. Shown in Figure \ref{fig:LSSwdens} is an example 
of a density map plotted for the protostructure in the CFHTLS-D1 field with the highest significance in photometric redshift galaxy density. 

\subsection{Discovery of a $z\sim3.3$ protostructure in the CFHTLS-D1 field}
\label{protocluster}

Of the 13 spectroscopically detected protostructures in the CFHTLS-D1 field using the search algorithm described above, one, 
a protostructure at $z\sim3.3$, far exceeded the others both in terms of the density of spectroscopic member galaxies and the 
density of potential photometric redshift members. As shown in Figure 
\ref{fig:sigLSS} and in Figures \ref{fig:LSS} and \ref{fig:LSSwdens}, this protostructure is detected extremely significantly
both in the number of spectroscopically confirmed member galaxies, $\delta_{gal}=10.5\pm2.8$, and in its overdensity of sources 
with photometric redshifts consistent with the protostructure redshift, $\sigma_{SEx,LSS}=8.0$ (see Figure \ref{fig:sigLSS} for the
meanings of these terms). While the nominal transverse size of our overdensity search was $R_{proj}<2$ $h_{70}^{-1}$ proper Mpc, in 
order to be as inclusive as possible while still probing a reasonably small volume, we allowed the defined transverse extent of the 
protostructure increase to the (projected) radius at which the galaxy density fell to $\sim50$\% of the density calculated with the nominal filter size. 

For the protostructure that is the subject of this paper, referred to hereafter as Cl J0227-0421\footnote{While a prefix designating this protostructure 
a cluster may seem presumptuous, the reason for this is formally quantified in \S\ref{halomass}.}, the projected radius at which
the galaxy density fell to this value was found to be $R_{proj}<3$ $h_{70}^{-1}$ proper Mpc. This distance 
is still easily spanned by $z=0$ for galaxies with transverse velocities in excess of even a small fraction of those of typical
low-redshift cluster galaxies. A similar size increase was not applied to the dimension of the filter along the line of sight since the size 
in this dimension already far exceeded that of the radial dimension, and, furthermore, a galaxy lying at further distances along the line of sight, 
subject to the assumption of radial infall at $\sim1000$ km s$^{-1}$, could not reach the core of the protostructure by $z=0$. This radial cut 
is used for all subsequent analysis with one exception mentioned later, though we note that all results for this protostructure,
including the magnitude of the spectroscopic overdensity, are largely insensitive to the specific choice of the size of the dimensions probed
for $R_{proj}<4$ $h_{70}^{-1}$ proper Mpc and $\Delta\chi<25$ $h_{70}^{-1}$ proper Mpc. We also tested for effects on 
$\delta_{gal}$ as a result of non-uniform spectral sampling and found no difference in the calculated value if the ``field'' search
described in Figure \ref{fig:sigLSS} was instead limited to the area over which the protostructure extended (i.e., the same VIMOS quadrant).

\begin{figure*}
\epsscale{1}
\plottwoalmostspecial{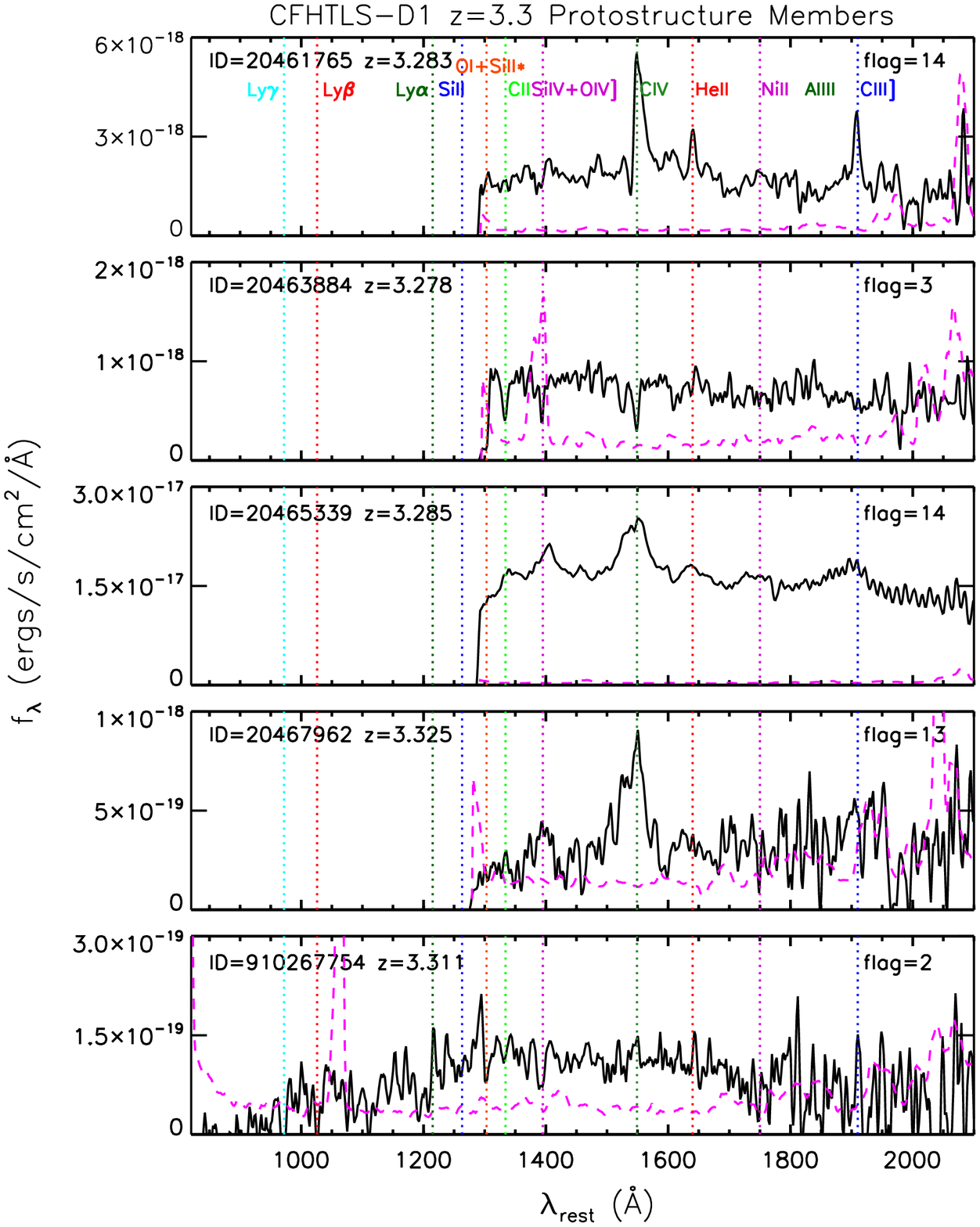}{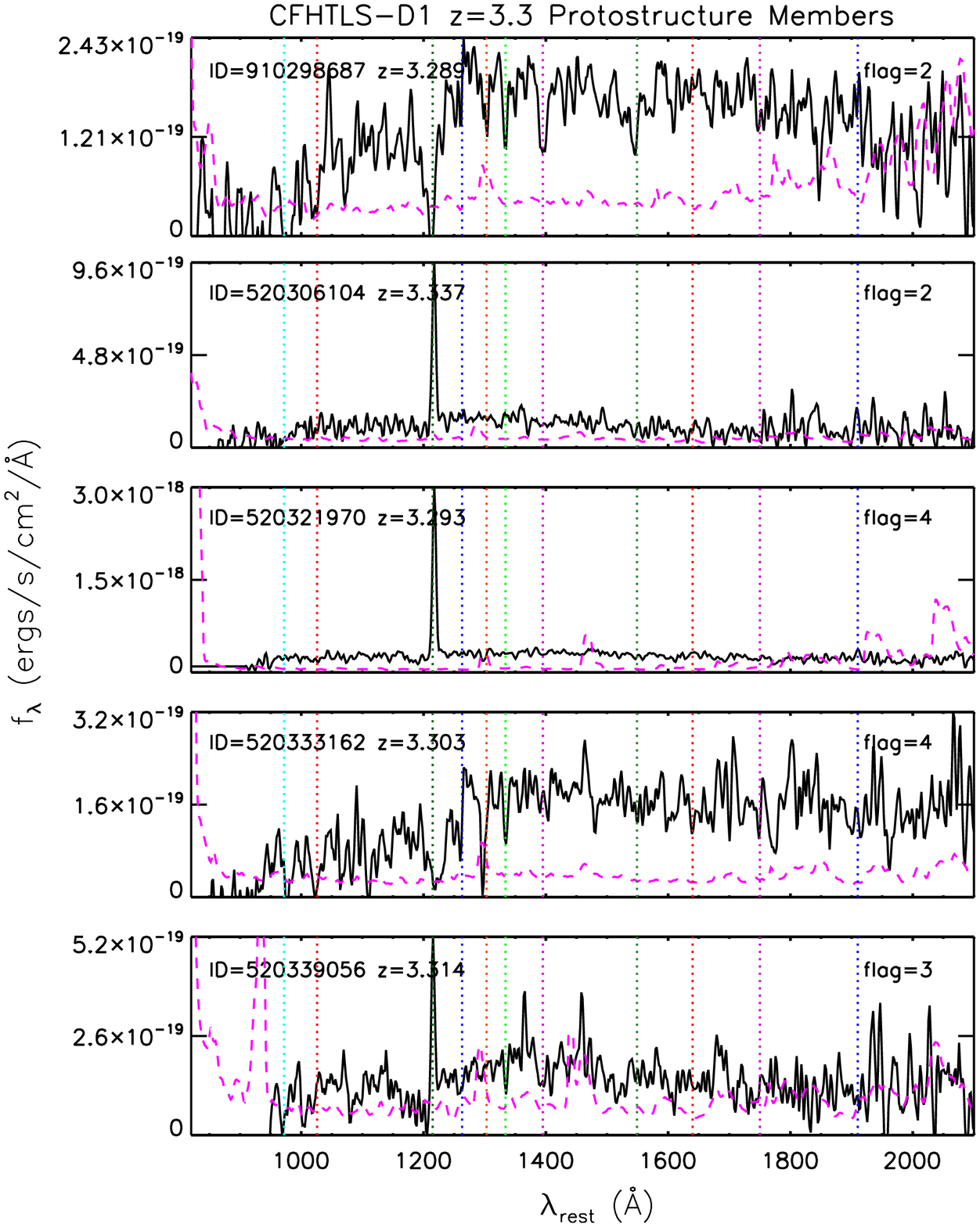}
\caption{Mosaic of the one-dimensional rest-frame VIMOS spectra of ten spectral members of Cl J0227-0421.
The black line in each panel is the flux density spectrum, and the dashed magenta line is the formal uncertainty spectrum
(see Le F{\`e}vre et al. 2014 and references therein for details on the generation of the uncertainty spectrum). Important spectral features
are marked. The spectrum of the proto-BCG, a type-1, and X-ray AGN host is shown
in the 3rd panel from the top on the left. The spectra of the two other type-1 AGN hosts are shown in the top and 4th from the
top panel on the left. The first four galaxies plotted in the left panel were observed as part of the VVDS-Deep sample and, as such, do not
have observed spectra blueward of $\lambda_{rest}\lsim1290$\AA. A wide range of diversity in spectral properties is seen among the protostructure members.}
\label{fig:specmosaic1}
\end{figure*}

\begin{figure*}
\epsscale{1}
\plottwoalmostspecial{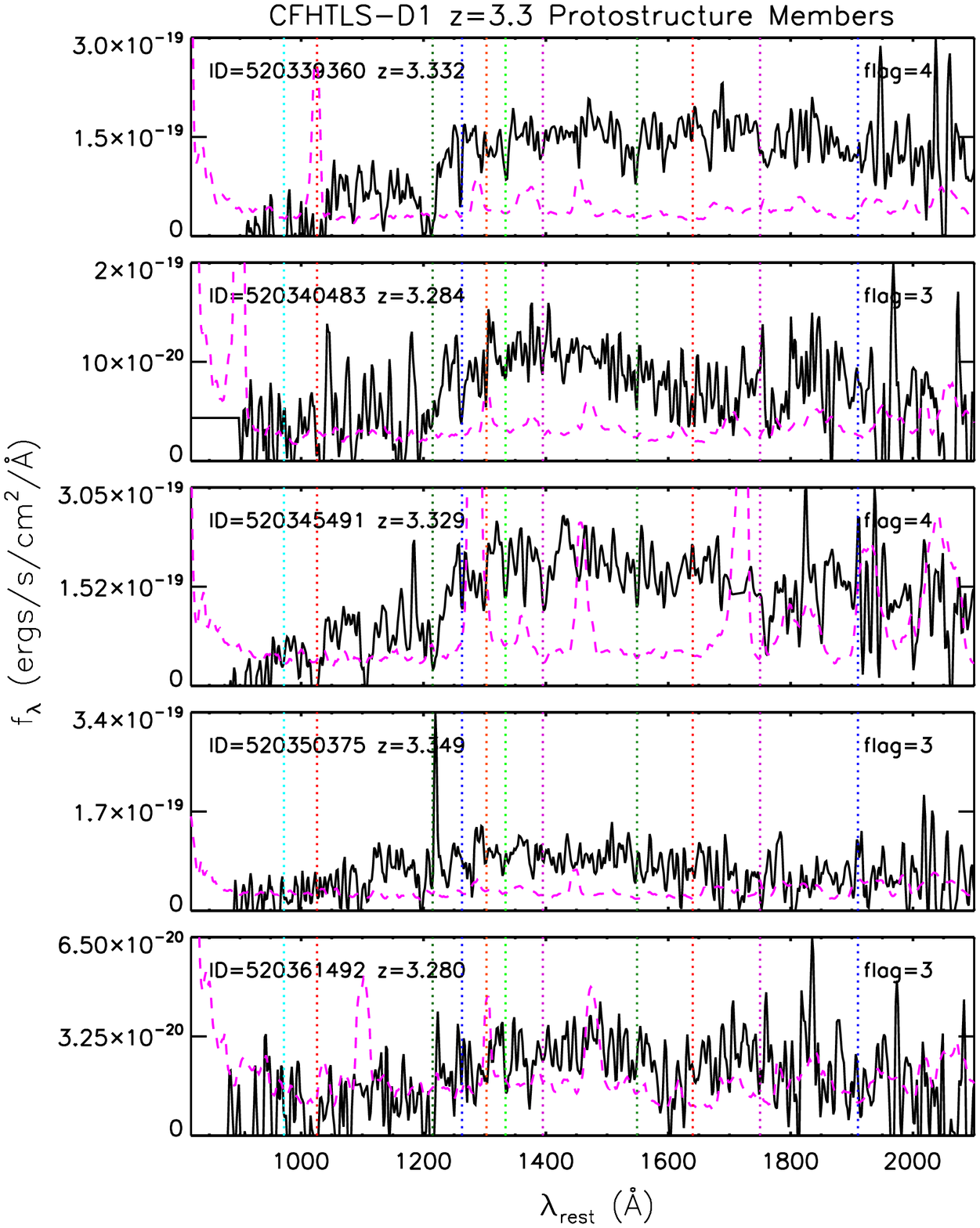}{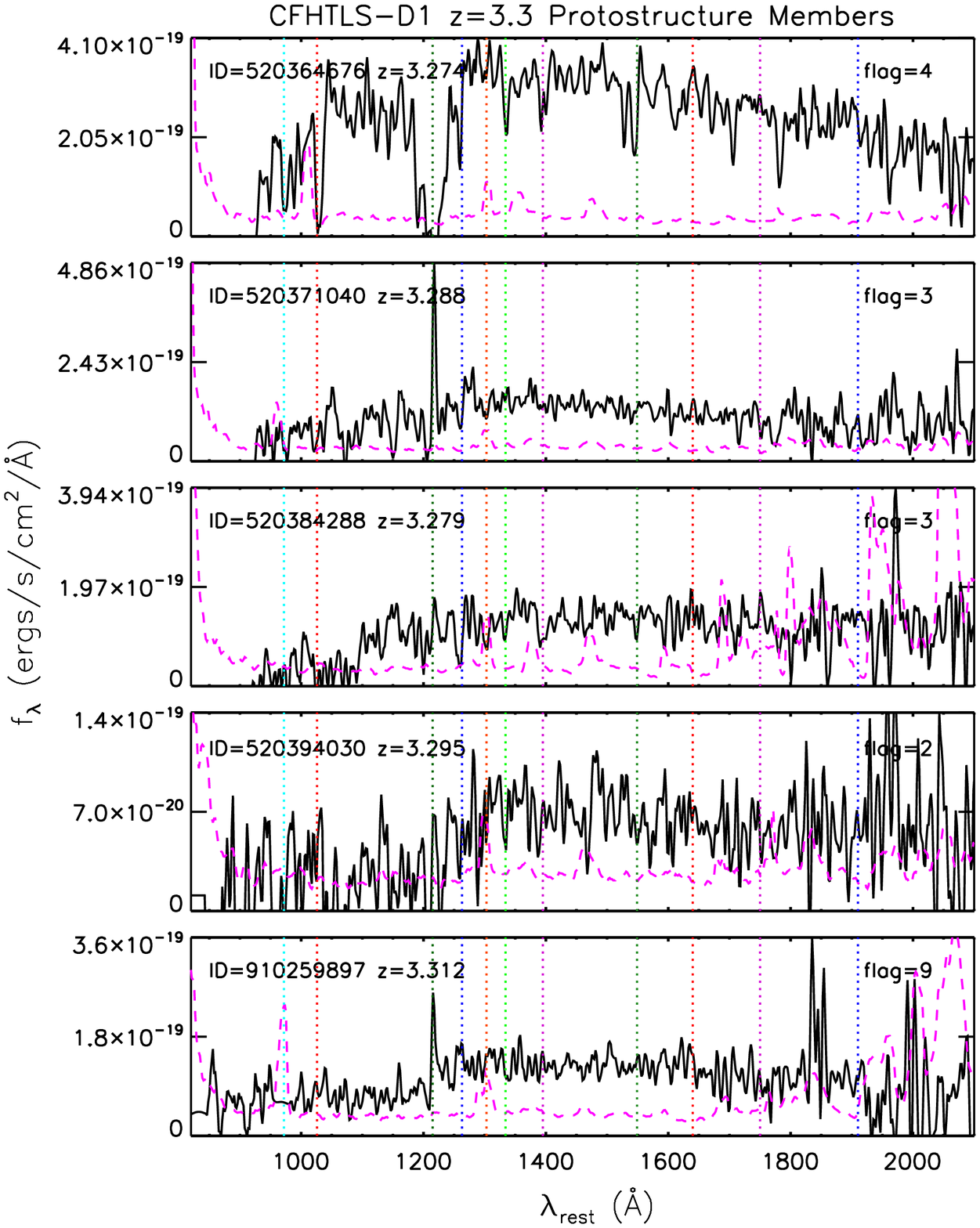}
\caption{Mosaic of the one-dimensional rest-frame VIMOS spectra remaining nine spectral members of Cl J0227-0421.
Also plotted in the bottom right hand panel (ID=910259897) is the only member with a less secure spectroscopic redshift that exhibits a strong
emission line in its spectrum. In this case we presume the line to be Ly$\alpha$, though this galaxy does not enter any of our analysis and is
presented here and elsewhere only for illustrative purposes. The meanings of all lines are the same as in Figure \ref{fig:specmosaic1}.}
\label{fig:specmosaic2}
\end{figure*}

The spatial center of Cl J0227-0421 was calculated in a method similar to the one described in Ascaso et al.\ (2014) for all galaxies 
within $3.27 < z < 3.35$ and $R_{proj}<3$ $h_{70}^{-1}$ Mpc, but with the peak of the photometric redshift source density map 
serving as the initial guess as the center. Unit weighting was chosen over luminosity weighting owing to significant 
contamination from AGN activity of the brightest galaxy in the protostructure (see \S\ref{protoBCG}) in both the $K_{s}$ 
and the IRAC bands. Regardless, the centers calculated from $K_{s}$-band luminosity-weighted average or a unit-weighted
average of members within $R_{proj}<2$ $h_{70}^{-1}$ Mpc are shifted negligibly from the adopted center ($\sim15\arcsec$ 
or $\sim100$ kpc at $z=3.3$), which if used instead, would have no effect on our results. In Figure \ref{fig:LSSwdens} a 
spectroscopic redshift histogram is plotted of all galaxies with $2.9<z_{spec}<4.0$ within 
$R_{proj}<3$ $h_{70}^{-1}$ of the number-weighted center. Both the 
unit-weighted spectroscopic center and the photometric member density center are given in Table \ref{tab:PSprop}, along 
with the number of members within the adopted bounds of Cl J0227-0421 and their median redshift. In total, 19 members with 
\emph{secure spectroscopic redshifts} are found within Cl J0227-0421 (referred to hereafter as ``spectral members''), 
with another six galaxies having spectroscopic redshifts consistent with that of the protostructure 
but with a lower reliability (i.e., flags=1 \& 9, referred to hereafter as ``questionable spectral members''). The latter galaxies are included
throughout the paper for illustrative purposes only and do not enter into any of our analysis except as potential photometric
redshift members. Figures \ref{fig:specmosaic1} and \ref{fig:specmosaic2} show the rest-frame VIMOS spectra of the 19 spectral members of 
Cl J0227-0421, along with the one questionable spectral member whose spectrum contains a single strong emission feature,
presumed in this case to be Ly$\alpha$. In Table 2, available through CDS, we give the identification number, right ascension,
declination, spectroscopic and photometric redshift, apparent and absolute magnitude, stellar mass, and SFR of each of the 
spectral members and questionable spectral members of Cl J0227-0421.

\begin{table*}
\caption{General properties of Cl J0227-0421\label{tab:PSprop}}
\centering
\begin{tabular}{cc}
\hline \hline
Spectral-number-weighted center & $[\alpha_{J2000},\delta_{J2000}]=[$02:27:00.6, -04:21:20.2$]$ \\[4pt]
Photo-$z$ density map center & $[\alpha_{J2000},\delta_{J2000}]=[$02:26:55.2, -04:20:45.6$]$ \\[4pt]
Number of spectral members & 19 (6)\tablefootmark{a} \\ [4pt]
Median redshift & $\tilde{z}=3.293$ \\[4pt]
Spectral overdensity & $\delta_{gal}=10.5\pm2.8$, $\sigma_{Proto-Struct.}=13.5$\\[4pt]
Photo-$z$ overdensity & $\sigma_{SEx,LSS}=8.0$\tablefootmark{b}\\[4pt]
Galaxy velocity dispersion & $\sigma_{v}=995\pm343$ km s$^{-1}$\\[4pt]
\hline
\end{tabular}
\tablefoot{
\tablefoottext{a}{The first number refers to all spectroscopically confirmed members with $R_{proj}<3$ $h_{70}^{-1}$ Mpc and $3.27<z_{spec}<3.35$. The number in
parentheses refers to tentative members with less reliable redshift measurements (see \S\ref{spectra}).}
\tablefoottext{b}{This number refers to the formal significance of the detection after accounting for spurious density peaks (see \S\ref{protocluster}).}
}
\end{table*}

\begin{figure}
\epsscale{1}
\plotone{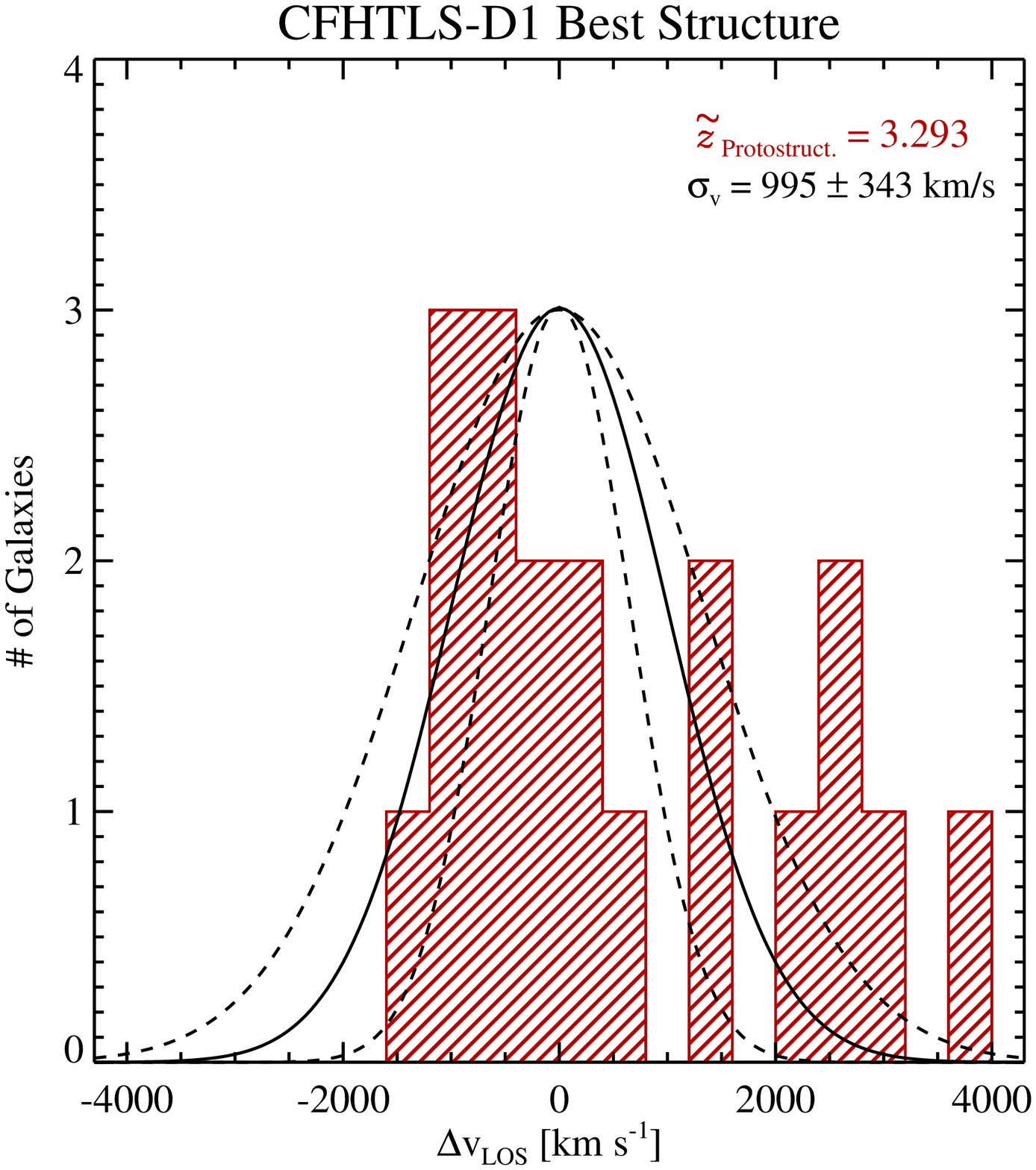}
\caption{Differential velocity distribution of the all spectral members of Cl J0227-0421. The median
redshift of the secure members is shown in the top right corner of the plot. Also shown in the top right corner is the value of the best-fit 
line of sight (LOS) velocity dispersion ($\sigma_{v}$, see \S\ref{protocluster} for details). The resulting Gaussian function generated by the 
best-fit $\sigma_{v}$ is overplotted on the differential velocity histogram (solid black line) along with those functions generated from 
$\sigma_{v}\pm\sigma_{\sigma_v}$. The high degree of skewness of the differential velocity distribution of member galaxies can be clearly seen.}
\label{fig:diffvel}
\end{figure}

With the relatively large number of spectral members afforded by the VUDS and VVDS spectroscopy, an attempt can 
be made to calculate the dynamics of the members of Cl J0227-0421. 
Because the member galaxies of this protostructure have had little time to interact, it is likely that the dynamics will
depart appreciably from the near-virialized dynamics observed in members of lower redshift structures. In 
addition, the estimated dynamics of cluster and group members that have had much more time to mature have been found to 
vary considerably with spectroscopic sampling, both in the number of members and the representative number of sampled 
galaxies of various types, making any estimate here highly uncertain. With this warning, the line-of-sight velocity dispersion (referred to
simply as the velocity dispersion hereafter), 
$\sigma_{v}$, was calculated using the method of Rumbaugh et al.\ (2013) for the 14 spectral members
within $R_{proj}<2$ $h_{70}^{-1}$ Mpc. A smaller radial cut was used here to probe galaxies that have had
a greater chance to interact with each other and the protostructure potential, though $\sigma_{v}$ does not vary within 
the errors if we instead chose to use all members within $R_{proj}<3$ $h_{70}^{-1}$ Mpc. 

Four different methods were used to 
calculate the velocity dispersion, identical to those of Rumbaugh et al.\ (2013), with errors estimated through jackknifing. 
The velocity dispersion estimated by the f-pseudosigma method, a method that performs adequately in probing the true distribution of
sparsely sampled non-Gaussian distributions (see Beers et al.\ 1990), was adopted as the best-fit velocity dispersion. Our results are not
heavily reliant on this choice because all other methods had values consistent with this values within 1$\sigma$ of their (large) formal errors. The differential
velocity distribution of member galaxies, plotted in Figure \ref{fig:diffvel}, is highly non-Gaussian, with a skewness of 1.06, probably a result of the (relatively) small 
number of member galaxies with secure spectroscopic redshifts and the high redshift of the protostructure. The f-pseudosigma galaxy 
velocity dispersion, which is the value used throughout the remainder of this paper, was calculated to 
be $\sigma_{v}=995\pm343$ km s$^{-1}$. The corresponding virial radius at the redshift of Cl J0227-0421, a quantity used extensively 
in the next section, was calculated using the methodology of Lemaux et al.\ (2012) to be $R_{vir}=0.46\pm$0.16 $h_{70}^{-1}$ Mpc. While we adopt this
value of the virial radius for the remainder of the paper, we do not require it for any of our analysis to have a physical meaning outside of a 
distance from the center of the protostructure, which represents some density contrast to which we can scale global quantities. Indeed, it has 
been suggested that a high percentage of the mass of a structure at a given redshift lies not within the virial radius at the redshift of a source, but
rather in the virial radius as estimated from the critical density evaluated at $z=0$ (Zemp 2013). The value of this quantity,
$R_{vir, \, z=0}=2.2\pm$0.7 $h_{70}^{-1}$ Mpc, is far more well matched to our protostructure filter and the criterion
used to define membership throughout this paper. The choice of adopting $R_{vir}$ at the redshift of the protostucture for use in our analysis 
was governed simply by convention and convenience, and another value, such as $R_{200}$ or $R_{vir, \, z=0}$, could have been chosen with no effect on our 
results. 

\subsubsection{The halo mass of Cl J0227-0421 and its predicted fate}
\label{halomass}

At lower redshift ($z\la1$) strong correlations are observed between the properties of cluster and group galaxies and the 
total mass of the structure in which they reside. The maturity of the dynamical evolution of a host structure or a 
perturbing event, such as a cluster-cluster merger, can also govern the properties of its galaxy content to some
degree 
(e.g., Ma et al.\ 2010; Lemaux et al.\ 2012; Rumbaugh et al.\ 2012; Stroe et al.\ 2014, though see also De Propris et al.\ 2013). However,
averaged over many structures, the halo mass has been found to be intimately linked to the fraction of blue, star-forming, 
and starbursting member galaxies, the properties of the brightest cluster and group galaxies, and the shape of member galaxy 
luminosity/stellar mass functions. While
still difficult
to measure and correctly calibrate, halo mass proxies at these redshifts are relatively numerous. The dynamics of large numbers of spectroscopically confirmed 
members galaxies, weak or strong gravitational lensing, and measurements of the properties of the hot intracluster 
medium, either through Bremsstrahlung emission or via the inverse-Compton scattering of cosmic microwave background photons 
have all been used effectively at $z\la1$ to measure the masses of galaxy clusters, and, to a lesser extent, galaxy groups.
Each methodology, however, loses effectiveness (in different ways) as the redshift of the observed structure increases, and indeed, 
few halo mass measurements, calculated via these methods, exist for structures with redshifts in excess of $z\ga1.5$. 

With a redshift of $z\sim3.3$, estimating the halo mass of Cl J0227-0421 is daunting. Because of the large uncertainties and 
large number of assumptions that are required of any particular method, in this section we attempt four different methods of estimating 
or constraining the halo mass Cl J0227-0421. In this section we briefly describe the methods used and the halo mass and
associated uncertainties that result from each line of reasoning. For further details on the framework, assumptions, and details
of each method see Appendix B. While the results of this exercise can be used to test the standard $\Lambda$CDM concordance 
model of cosmology, as done in numerous other works investigating high-redshift structures (e.g., Foley et al.\ 2011; Gonzalez et al.\ 2012; Bayliss et al.\ 2013), 
the goal here is to simply provide a greater context for Cl J0227-0421 with which to compare other high-redshift protostructures and to provide a 
backdrop for the preliminary investigation of galaxy evolution that follows. 

We begin by calculating the halo mass of Cl J0227-0421 from the dynamics of the spectral members. The calculation was performed in a method identical
to Lemaux et al.\ (2012), though the impact of adopting assumptions valid at $z\sim1$ for a forming structure at $z\sim3.3$
are discussed in Appendix B. Using the value of the velocity dispersion from the previous section yields 
$\mathcal{M}_{dyn, \, vir}=3.16\pm2.18\times10^{14}$ $h_{70}^{-1}$ $M_{\odot}$. Given that
such high mass appears to already be in place at such high redshift, it is interesting to consider what the potential evolution of the halo of Cl J0227-0421  
would be to the present day. Adopting the formalism of McBride et al.\ (2009) and Fakhouri \& Ma (2010) based on results from the Millennium and 
Millennium-II simulations, the mean halo growth rate as a function of redshift and halo mass is defined as

\begin{multline}
\langle \dot{\mathcal{M}} \rangle_{mean} = 46.1 \, M_{\odot} \, \text{yr}^{-1} \left(\frac{\mathcal{M}_z}{10^{12} M_{\odot}}\right)^{1.1} \, (1+1.11z) \\
\times\sqrt{\Omega_{m,0}(1+z)^3 + \Omega_{\Lambda,0}}
\label{eqn:meangrowth}
\end{multline}

\noindent where $\mathcal{M}_z$ is the halo mass of the protostructure at the redshift of interest. Using the dynamical mass calculated above, integrating
this formula from $z=0$ to the adopted systemic redshift of Cl J0227-0421 ($z=3.29$), multiplying by the difference in the age of the universe at the two redshifts,
and adding the derived mass of Cl J0227-0421 at $z=3.29$ yields $\mathcal{M}_{dyn,\, vir, \, z=0}=8.69\pm4.71\times10^{15}$ $h_{70}^{-1}$ $M_{\odot}$. Errors are determined from those in the velocity dispersion. The halo mass estimated from this calculation is enormous, enough to rival the most massive galaxy clusters observed in the local universe 
(e.g., Piffaretti et al.\ 2011; Wang et al.\ 2014). However, the number of assumptions, their associated uncertainties, and the formal errors coming from the velocity dispersion calculation 
are also enormous. In addition, the above formula is meant to be applied to a single halo, whereas the dynamical mass estimate above may make use of galaxies
that populate several different halos, a subtlety that we have, with the current data, no power to constrain. If it is indeed the case that the galaxies 
used to estimate the dynamical mass of Cl J0227-0421 populate multiple halos, the z=0 mass estimated here will necessarily be an upper limit, though how constraining 
this limit depends on the multiplicity, mass ratio, and the proximity of the constituent subhalos. Regardless, such an effect is unlikely to be greater than the formal 
uncertainties in the evolved halo mass. It is sufficient to say, then, that the dynamical mass estimate places Cl J0227-0421 as a progenitor of a cluster within similar 
to or exceeding the mass of the Coma cluster ($\mathcal{M}_{dyn}\sim1-2\times10^{15}$ $M_{\odot}$; Kent \& Gunn 1996; Colless \& Dunn 1996).


A second approach is to use the stellar content of the protostructure as a proxy for the total mass. An estimate from this method 
is, however, likely to be a lower limit due to some fraction, perhaps considerable at these redshifts (see, e.g., Capak et al.\ 2011), of the 
baryonic content of member galaxies residing in unprocessed gas. Briefly, the calculation takes the form of summing up the total 
stellar mass content in all members of the protostructure within a certain radius, accounting for the missed number of members, 
and using the resulting total stellar mass of the members to estimate the total halo mass based on known correlations. For more
details, see Appendix B. The resulting halo mass is scaled to a common radius with that of all other methods (where we chose the virial radius 
for convenience) using a Navarro-Frenk-White (NFW, Navarro et al. 1996) profile as 
described in Appendix B. This halo mass estimate from this method was $M_{\Sigma\mathcal{M}_{s}, \, vir}= 1.87\pm0.98 \times10^{14}$ $h_{70}^{-1}$ $M_{\odot}$, 
consistent within approximately $1\sigma$ with the dynamical halo mass estimate. This halo mass was evolved to $z=0$ using the same methodology as 
for the dynamical mass results in a present-day halo mass of $M_{\Sigma\mathcal{M}_{s}, \, vir \, z=0} = 4.89\pm2.53 \times10^{15}$
$h_{70}^{-1}$ $M_{\odot}$.

\begin{table}
\renewcommand\thetable{3}
\caption{Halo mass estimates of Cl J0227-0421\label{tab:massprop}}
\centering
\begin{tabular}{lcc}
\hline \hline
Method & $\mathcal{M}_{vir, \, z=3.3}$\tablefootmark{a} & $\mathcal{M}_{vir, \, z=0}$\tablefootmark{a} \\[0.5pt]
\hline
Dynamics & $3.16\pm2.18\times10^{14}$ & $8.69\pm4.71\times10^{15}$ \\[4pt]
Stellar mass & $1.87\pm0.98 \times10^{14}$ & $4.89\pm2.53 \times10^{15}$ \\[4pt]
X-ray & $<3.35\pm1.46\times10^{14}$ & --- \\[4pt]
Galaxy density & --- & $3.67^{+1.55}_{-1.41}\times10^{15}$\\
\hline
\end{tabular}
\tablefoot{
\tablefoottext{a}{In units of $h_{70}^{-1}$ $M_{\odot}$ evaluated at the virial radius of Cl J0227-0421 ($R_{vir}=0.46$ $h_{70}^{-1}$ Mpc)}
}
\end{table}

As mentioned previously, the CFHTLS-D1 field was imaged with \emph{XMM-Newton}/EPIC to a depth of 10.6 ks in the proximity of 
Cl J0227-0421\footnote{Another pointing of XMM was centered to the northeast of the protostructure to a depth of 24 ks, but the large off-axis angle of 
the protostructure in this pointing resulted in an X-ray flux limit that was similar to that of the 10.6 ks exposure.}
While this depth is not sufficient to significantly detect X-ray emission from any nascent ICM that may exist in the protostructure, we 
determined an upper limit on this emission of $f_{X,\, [0.5-2] \, \rm{keV}}<1.29\pm0.31\times10^{-14}$ ergs s$^{-1}$ cm$^{-2}$ by the method described in Appendix B. This flux limit was converted 
into an observed-frame luminosity value at the redshift of Cl J0227-0421 and $k$-corrected with the Chandra Interactive Analysis of Observations package (CIAO; Fruscione et al.\ 2006) 
to the rest-frame [0.1-2.4] keV band using a Raymond-Smith thermal plasma model (Raymond \& Smith 1977) with a temperature of 2 keV and an abundance of 0.3$Z_{\odot}$ (though using models 
of differing temperatures or abundances gives consistent results within $\sim50$\%). This luminosity
limit was in turn used to estimate a hydrostatic mass limit within $r_{500}$ and was transformed to a mass limit at the virial radius using the methods described in Appendix B.
The resulting hydrostatic halo mass limit is $M_{X,\, vir}<3.35\pm1.46\times10^{14}$ $h_{70}^{-1}$ $M_{\odot}$. Because this value is a limit, we do not attempt to evolve it to the 
present day.

The final halo mass calculation is based on translating the spectroscopic overdensity into a halo mass using a relationship between the clustering of galaxies 
and their underlying dark matter distribution. This methodology relies heavily on the one presented in Chiang et al.\ (2013) and Steidel et al.\ (1998), and the manifestation
of this methodology that was adapted for this work is presented in Cucciati et al.\ (2014). As such, we only mention those aspects relevant for this calculation on
Cl J0227-0421 and refer interested readers to those studies. The galaxy overdensity, $\delta_{gal}$, calculated in \S\ref{protocluster} was calculated again using a box filter with 
half-height dimensions of $R_{e}=8.5$ comoving Mpc, appropriate for a protostructure at $z\sim3.3$, yielding $\delta_{gal}=13.3\pm6.6$. For this calculation 
a stellar mass limit of $\mathcal{M}_{s}>10^9$ $M_{\odot}$ was imposed on both the spectral members, and the field and a galaxy bias, $b=2.38$, was adopted based on 
a linear interpolation of biases at different redshifts presented for an identical stellar mass cut in Chiang et al.\ (2013). 

At this point a long overdue matter of nomenclature needs to be mentioned regarding the designation of Cl J0227-0421. Having now calculated $\delta_{gal}$ for an equivalent 
sample as presented in Chiang et al.\ (2013), we can \emph{directly} compare this value to the simulated protostructures from Chiang et al.\ (2013) to estimate the probability 
that Cl J0227-0421 will evolve to a cluster by $z=0$. Even the 1$\sigma$ lower bound of $\delta_{gal}$ calculated for Cl J0227-0421 exceeds the threshold at which a protostructure 
will \emph{always} evolve into a cluster as determined for an identical filter size at an identical redshift in the Millennium simulations. In this way we justify the designation
of Cl J0227-0421 as a cluster in the process of formation thus allowing us to refer to it as a protocluster for the remainder of the paper. From the calculated $\delta_{gal}$ and 
adopted bias factor the halo mass of Cl J0227-0421, evolved to $z=0$ from the calibrations in Chiang et al.\ (2013), was estimated following the methodology outlined in Appendix B. As with all other 
estimates, the halo mass estimate was transformed to that at the virial radius resulting in a value of $\mathcal{M}_{\delta_{m}, \, vir, \, z=0}=3.67^{+1.55}_{-1.41}\times10^{15}$ $h_{70}^{-1}$ $M_{\odot}$.
The constraints on the halo mass of the Cl J0227-0421 protocluster placed by all four methodologies are summarized in Table \ref{tab:massprop}.

Though extremely large uncertainties exist both formally and in the assumptions made to derive the values given in Table \ref{tab:massprop}, and perhaps because of this, the
high degree of concordance between the values derived from four methods is astonishing. While the exact value of the halo mass of Cl J0227-0421 can only be constrained, at best, within
a factor of $\sim3$, the values given in Table \ref{tab:massprop}, along with the high value of $\delta_{gal}$ presented in both this section and \S\ref{analysis}, paint the 
picture that Cl J0227-0421 is a protocluster with a large amount of mass already assembled very early in the history of the universe and that it is 
destined to descend into a cluster whose mass will rival or exceed the Coma cluster. With this global picture in mind, we proceed to make a 
preliminary investigation into the properties of the galaxies housed within this emerging cluster. 

\begin{figure*}
\epsscale{0.7}
\plottwonotsospecial{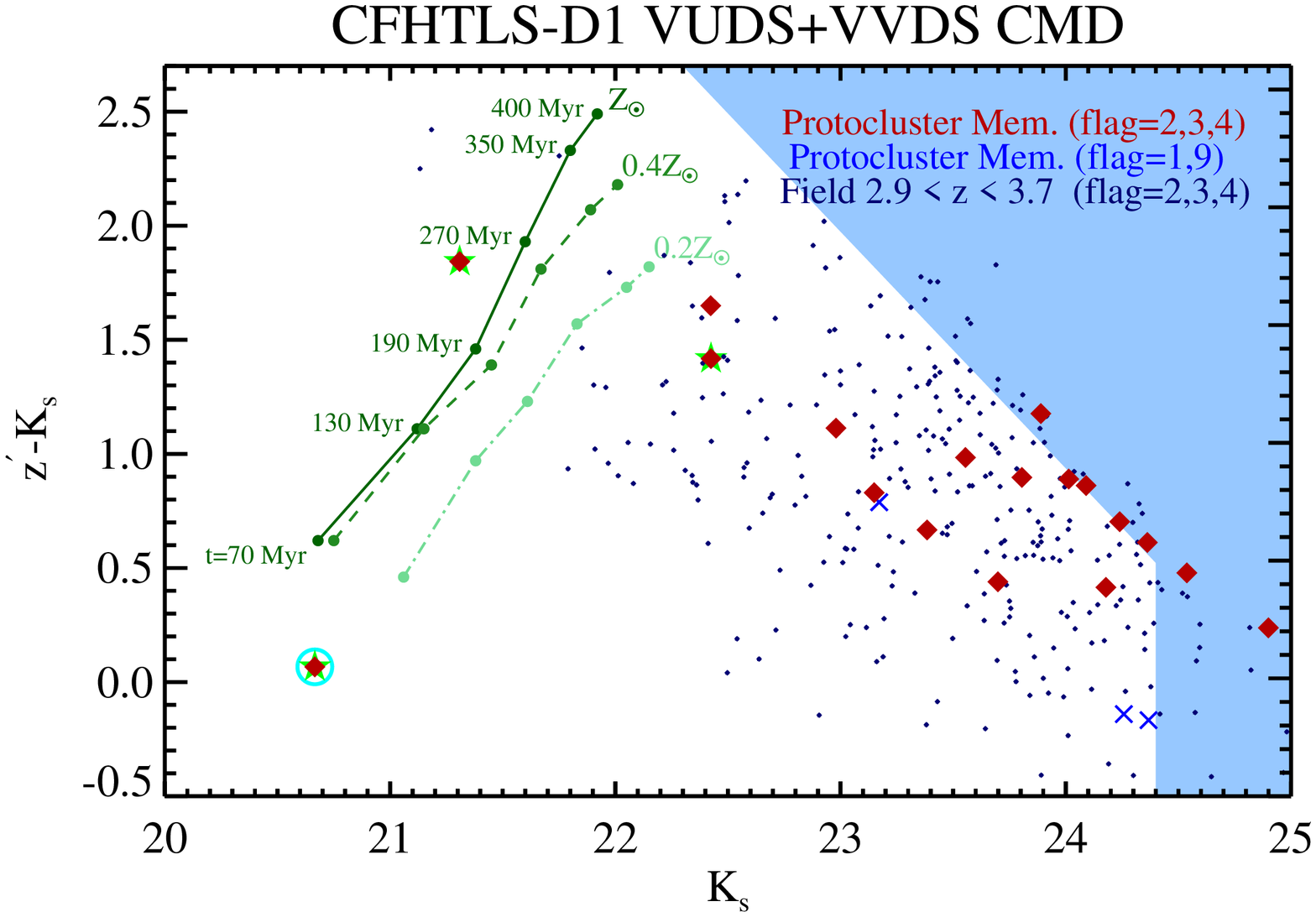}{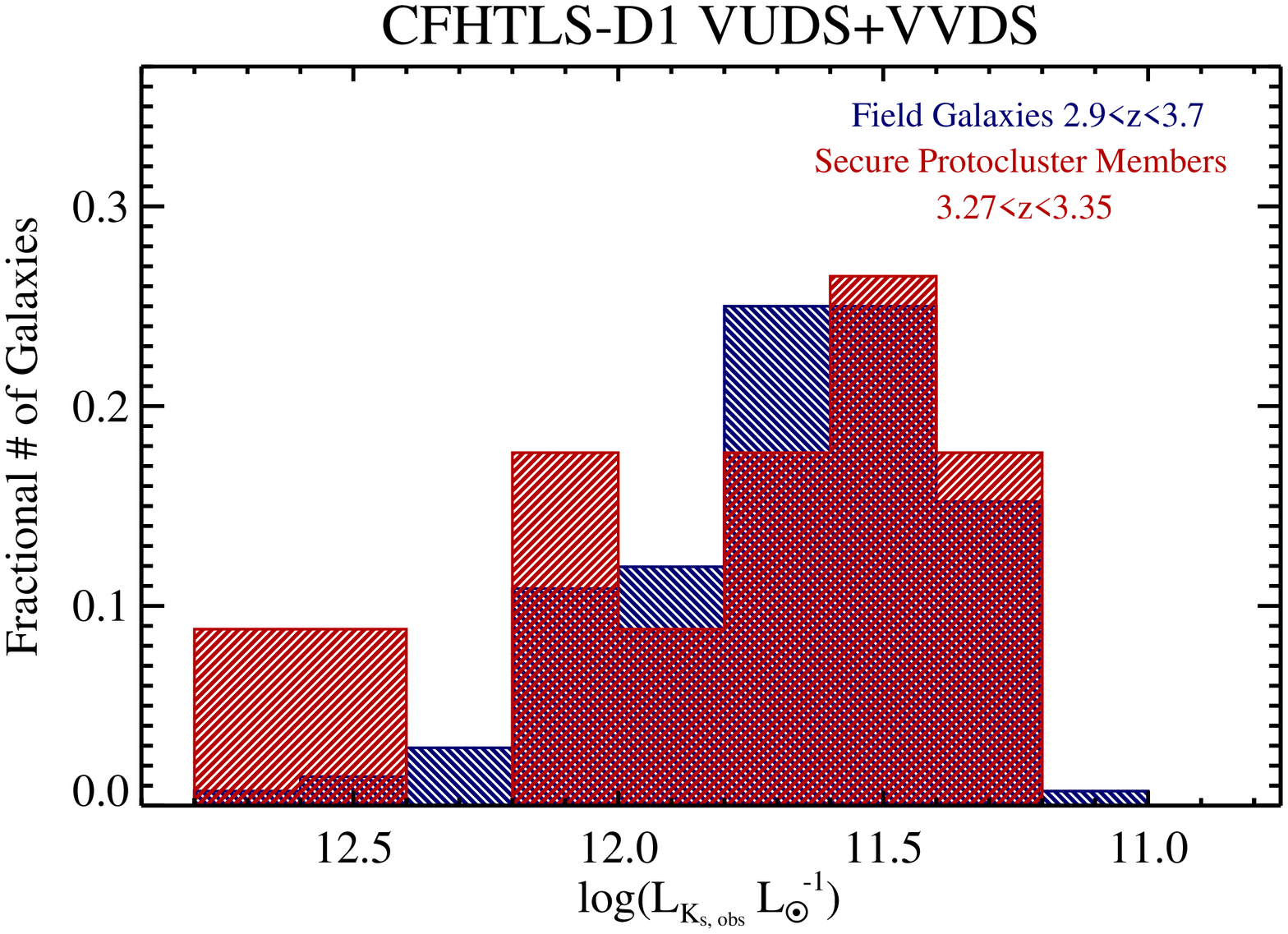}
\caption{\emph{Left:} Observed-frame CFHTLS/WIRDS color-magnitude diagram (CMD). Members of Cl J0227-0421 with secure spectroscopic redshifts are
shown as red diamonds, those with less secure spectroscopic redshifts are shown as blue Xs. The meanings of the green star and cyan circle are the same
as those in Figure \ref{fig:LSS}. All galaxies in the entire CFHTLS-D1 field with secure spectroscopic redshifts $2.9<z_{spec}<3.7$ that do not reside in the bounds of Cl J0227-0421
(i.e., ``field galaxies'', see \S\ref{CMDnCSMD}) are plotted as small navy points. The light blue-shaded region indicates the region of this phase space                                                            
not probed to the 5$\sigma$ point-source completeness limits of the CFHTLS/WIRDS imaging. Model tracks for a $z=3.3$ $L^{\ast}$ galaxy with three different stellar         
metallicities are overplotted (see \S\ref{protoRSG}). Several galaxies which are extremely bright in the $K_{s}$ band, including two of the four brightest galaxies 
in the entire spectroscopic sample over this redshift range, are members of the protocluster. \emph{Right:} Fractional distribution of observed-frame $K_s$-band luminosities 
in the protocluster and field samples. Only those protocluster members with secure spectroscopic redshifts and only those galaxies lying outside of the blue shaded region in the
left panel are plotted. Though a large fraction of the protocluster member galaxies have relatively normal $K_s$-band luminosities with respect to the general field population, 
the percentage of bright ($\log(L_{K_s})\gsim12.0$) galaxies residing in Cl J0227-0421 is nearly double that of the field.}
\label{fig:obsCMD}
\end{figure*}

\section{The Effect of environment in Cl J0227-0421}
\label{memprop}

The Cl J0227-0421 protocluster is characterized by 19 confirmed members, six additional potential spectroscopic members, and a high density of spectroscopic sampling over the 
entire spatial and redshift extent of the protocluster. Despite this, the number of member galaxies remains small relative to samples at lower redshifts where questions about 
galaxy evolution still abound. Complicating matters, the dominant environmental process or processes appears to depend non-trivially on the 
particular structure or structures being observed, the spectroscopic sampling of member galaxies, and the data available, especially the presence or absence of 
multi-wavelength\footnote{By multiwavelength data we mean here and throughout the paper any imaging data blueward or redward of the typical optical/NIR imaging available 
for most environmental studies.} data (see, e.g., the review in Oemler et al.\ 2009). With a sample size of one, we can only hope to provide an initial and cursory glance at
the effect of environmental processes (or lack thereof) on galaxy evolution in the high-redshift universe by studying the galaxy population of Cl J0227-0421. 

Compounding 
the difficulty of this study is the high redshift of the protocluster. At the redshift of Cl J0227-0421 the bandpasses of our ground-based optical/NIR imaging, as well as 
our optical spectral coverage, have been pushed far to the blue in the rest frame. As a result, the spectral and photometric diagnostics typically employed for galaxy 
evolution studies are either of questionable accuracy or impossible with the current data. While the accuracy and possible limitations or biases to the SED-fitting process
are mentioned in Appendix A, we stress here that the testing of the SED-fitting process, as well as understanding the proper methods to extract relevant parameters and their 
associated uncertainties from the rest-frame, near-ultraviolet (NUV) spectra, is still an ongoing investigation in VUDS. With these warnings, we begin a preliminary
investigation into the effect of environment in the early universe, deferring more complex analysis to future work with the full VUDS sample.

\subsection{Color$-$magnitude and color$-$stellar-mass properties}
\label{CMDnCSMD}

Plotted in the left panel of Figure \ref{fig:obsCMD} is the observed-frame $z^{\prime}-K_{s}$ color$-$magnitude diagram (CMD) of the spectral members of Cl J0227-0421. These two bands were 
chosen because they bracket the Balmer/4000\AA\ break at the redshift of the protocluster. The $K_{s}$ band was preferred over either the [3.6] or the [4.5] magnitude because the
WIRDS imaging is marginally deeper than that of SERVS. Also plotted in the lefthand panel of Figure \ref{fig:obsCMD} are all galaxies with secure spectroscopic redshifts from $2.9<z<3.7$ 
not associated with an overdensity. This sample of $\sim$500 galaxies, referred to hereafter as ``field'' galaxies, was chosen to represent a control sample for the
spectral members of Cl J0227-0421 at roughly the same epoch\footnote{The field redshift window represents a $\sim\pm300$ Myr window roughly centered on the cosmic time 
measured at the protocluster redshift}. One of the most striking features of the observed-frame CMD is that, while the protocluster galaxies
are found in an extremely small volume relative to the full field galaxy sample (see \S\ref{protoRSG}), the galaxies in the two samples essentially span the same region of 
color$-$magnitude phase space. While a large number of the spectral members lie at rather ordinary magnitudes and colors with respect to the field population, several galaxies 
exist within the protocluster bounds that are extremely bright and exhibit (typically) redder observed-frame colors. As can be seen in the righthand panel of Figure \ref{fig:obsCMD}, 
where the fractional observed-frame $K_{s}$ luminosity\footnote{This luminosity was calculated using the $K$-band luminosity of the Sun 
(http://www.ucolick.org/$\sim$cnaw/sun.html) $k$-corrected in the observed-frame
to the CFHT WIRCam $K_s$ filter.} distribution of both samples is plotted, not only does Cl J0227-0421 contain several bright galaxies, but such galaxies also make a greater contribution
to the overall population than similar galaxies in the field. This difference is considerable, since the fraction of protocluster member galaxies with $\log(L_{K_s})\ga12.0$
is nearly double that of the field (33.0\% and 16.8\%, respectively). The properties of these bright, and typically redder protocluster galaxies, were foreshadowed in Figure \ref{fig:LSSwdens} 
and will be discussed extensively throughout this section.

\subsubsection{The brightest protocluster galaxy}
\label{protoBCG}

The one galaxy in this population that is an exception is the brightest galaxy in the protocluster, referred to hereafter as the ``proto-BCG''. This galaxy is extremely 
bright in the $K_{s}$ band ($K_{s}=20.67$), but exhibits extremely blue colors ($z^{\prime}-K_{s}=0.1$), the only galaxy from $2.9<z_{spec}<3.7$ in our sample that occupies
this region of phase space. The properties of this galaxy are worth discussing briefly. It has been well documented that high-density peaks or protoclusters
are more common in the regions surrounding high-redshift radio-loud quasars. However, to the 3$\sigma$ depth of our VLA 
data at $z\sim3.3$ ($P_{\nu,\, 1.4GHz,\, 3\sigma} < 25.4$ W Hz$^{-1}$)\footnote{$k-$corrections for X-ray and radio point sources were calculated following
the methods described in Lemaux et al.\ (2013).}, neither this object nor any other protocluster member is detected in the radio. This limit is much lower 
than the typical output of high-redshift, radio-loud quasars ($\log(P_{\nu,\, 1.4GHz})\ga27$ W Hz$^{-1}$), precluding the possibility that this galaxy contains an analogous 
phenomenon to those used in other large surveys as signposts for overdense environments (e.g., Wylezalek et al.\ 2013, 2014).

The rest-frame NUV spectrum of the proto-BCG does, however, contain several high-ionization emission features whose FWHMs are several 
1000 km s$^{-1}$, attesting to the presence of an active central engine. The proto-BCG is also the only spectral member to be detected in the \emph{XMM-Newton} imaging, and it 
has a rest-frame, full-band luminosity\footnote{The reported luminosity is corrected for Galactic absorption, see Chiappetti et al.\ (2005)} of 
$L_{X,[0.5-10 ,\ \rm{keV}]}=1.0\pm0.4\times10^{45}$ ergs s$^{-1}$, placing the AGN in the proto-BCG well within the QSO regime (e.g., George et al.\ 2000). 
Given the immense energy output of this AGN it is possible that it is either a progenitor or a descendant of the high-power, radio-loud quasars found in other overdense 
environments. The host galaxy is also the only spectral member to be even moderately detected in the \emph{Herschel}/SPIRE imaging. The formal significance of the 
detection is 2.5$\sigma$, which falls below the formal limit required for a secure detection. However, the proto-BCG is also detected, significantly, at 24$\mu m$, 
giving us some additional confidence that the SPIRE detection is legitimate. Tentatively assuming this detection is real, the total infrared luminosity of the proto-BCG implies that it is 
forming stars at a rate of $SFR_{proto-BCG}=750\pm70$ $M_{\odot}$ yr$^{-1}$. The proto-BCG is also located at a large (projected) distance from the protocluster center 
(1.1 $h_{70}^{-1}$ proper Mpc), a property that is typical in lower redshift clusters still undergoing formation (e.g., Katayama et al.\ 2003; Fassbender et al.\ 2011; 
Zitrin et al.\ 2012; Lidman et al.\ 2013). It appears that the proto-BCG of Cl J0227-0421 is still very much in the process of evolving. 

\begin{figure*}
\epsscale{1}
\plottwo{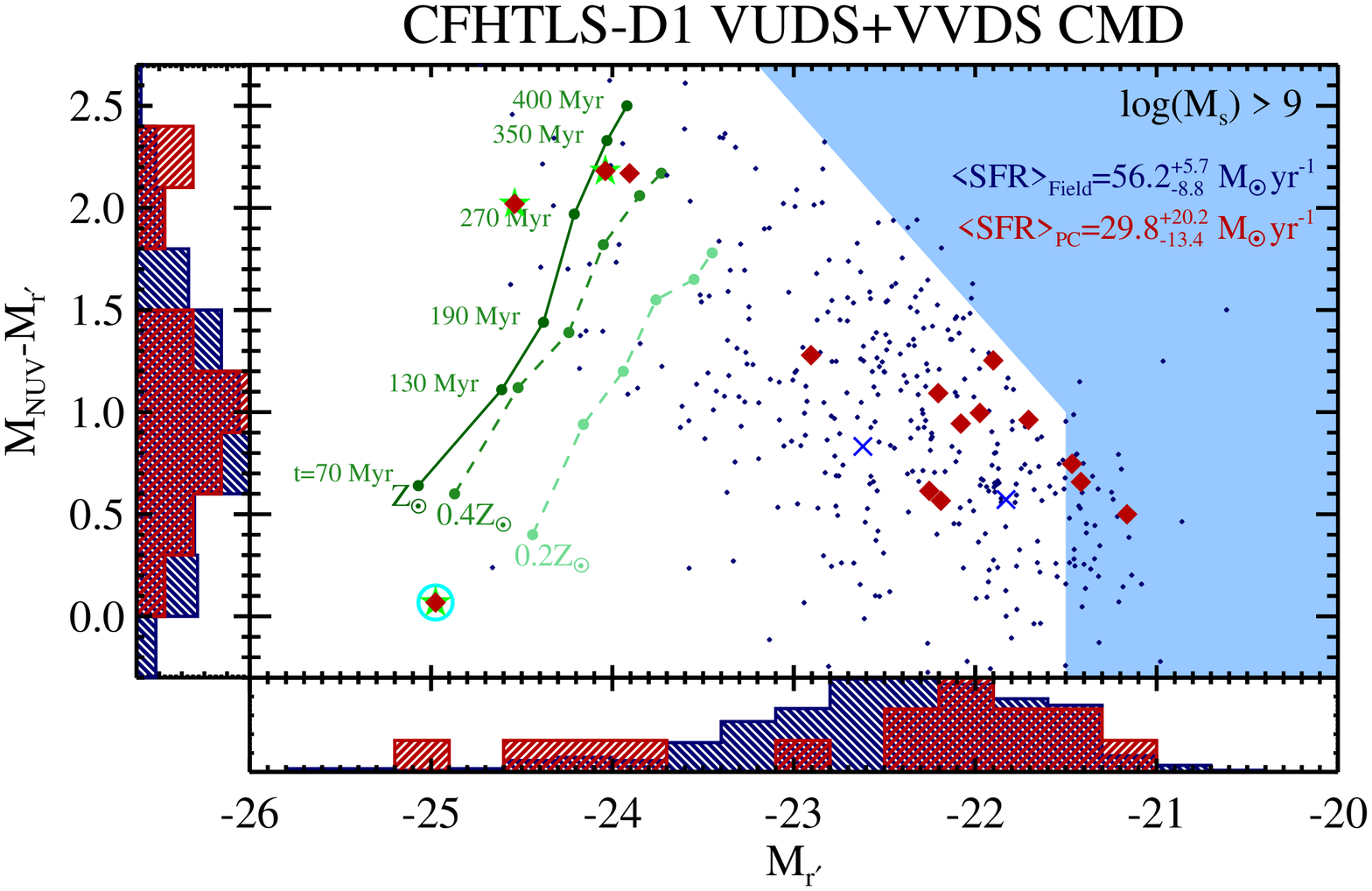}{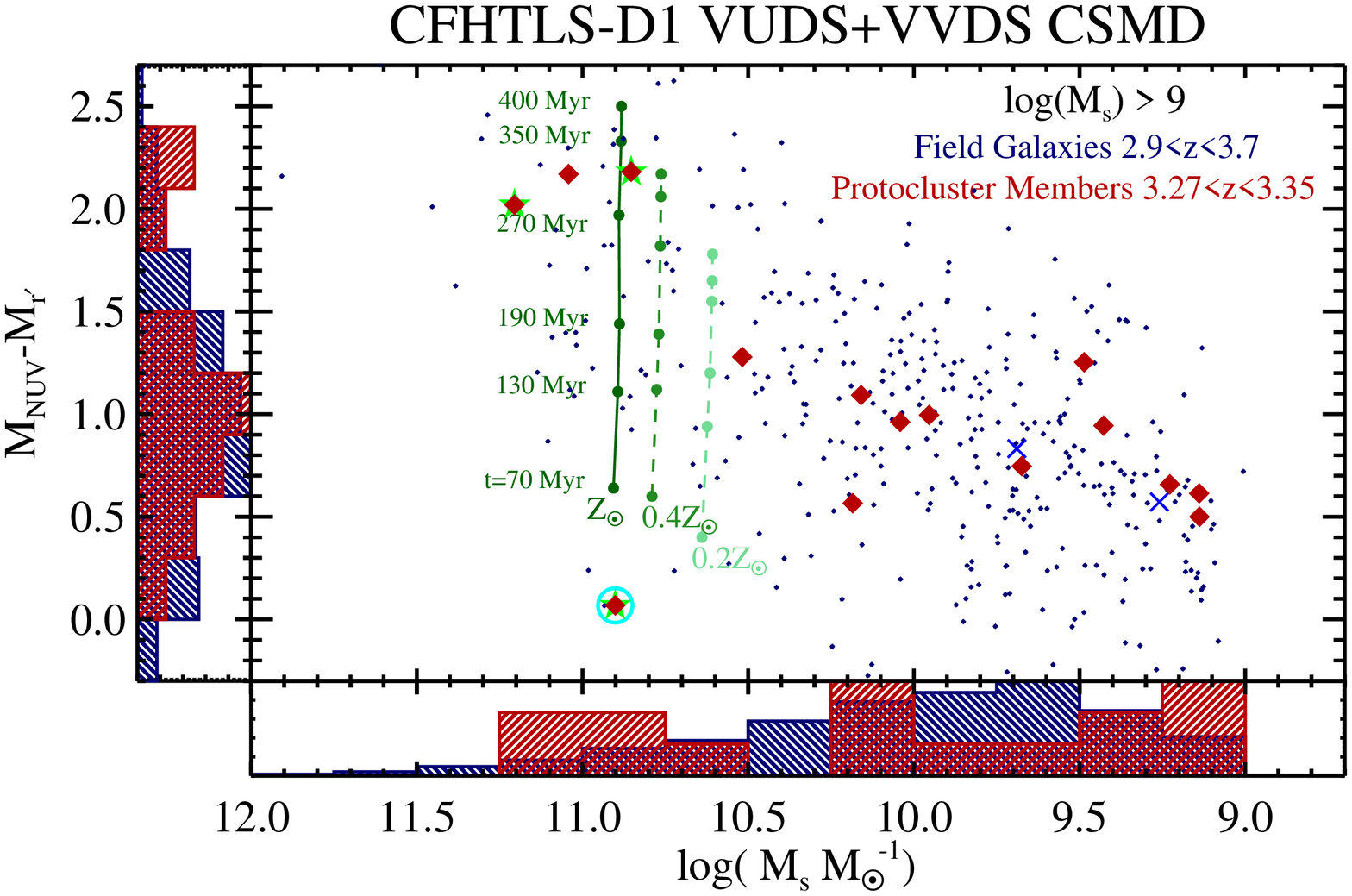}
\caption{\emph{Left:} Rest-frame $M_{r^{\prime}}$/$M_{NUV}-M_{r^{\prime}}$ CMD of all galaxies in the entire CFHTLS-D1 field with secure redshifts $2.9<z_{spec}<3.7$
and stellar masses $\log(\mathcal{M}_{s})>9$. The stellar mass cut is imposed here in an attempt to mitigate any induced differential bias between the field and 
protocluster members (see \S\ref{protoRSG}). The meanings of the symbols are identical to those of Figure \ref{fig:obsCMD} as is the meaning of the light blue-shaded region 
and the galaxy model tracks. Here and in the right panel, color and absolute magnitude histograms, normalized such that the maximum value is unity, are shown for each population.
While most protocluster members have relatively typical colors and 
magnitudes with respect to the field, there exists a sub-dominant population of extremely bright (and typically redder) protocluster galaxies. The median SFRs, as 
derived from the SED fitting process, of the two samples is shown in the upper right hand corner. \emph{Right:} Color-stellar-mass (CSMD) of the same galaxy populations 
shown in the left panel. The meanings of all symbols are the same. The bimodality observed in the CMD remains in the CSMD, with several protocluster galaxies having 
with extreme stellar masses $\log(\mathcal{M}_s) \gsim 10.8$. These galaxies comprise some of the most massive galaxies in the entire spectroscopic sample in the range 
$2.9<z_{spec}<3.7$. Though several field galaxies exhibit similar colors and stellar masses, the volume used to define the field sample is $\sim250$ larger than that used to define
the bounds of the protocluster. The proto-BCG, marked by the circumscribed cyan circle, likely has its stellar mass estimate contaminated by the presence of its 
powerful AGN, though the other two type-1 AGN hosts likely do not.}
\label{fig:restframeCMDnCSMD}
\end{figure*}

\subsubsection{The density of massive galaxies in Cl J0227-0421}
\label{protoRSG}

We now turn back to the bright galaxies in the protocluster that are observed at redder colors. At lower redshift, massive clusters, like the one that Cl J0227-0421 is predicted 
to evolve into, show marked increases in the abundances of bright and massive red-sequence galaxies (RSGs) relative to less dense environments 
(e.g., Ball et al.\ 2008; Wetzel et al.\ 2012). The origin of such galaxies is the subject of much debate, since it is unclear how early and through which processes 
such galaxies built up consider stellar masses and eventually quenched. In this respect, the presence of several bright galaxies already within the bounds of the protocluster 
at $z\sim3.3$ is tantalizing. However, in the early universe, especially given the relatively blue rest-frame wavelength coverage of the optical/NIR imaging employed here, it 
is far from certain that a direct connection can be drawn between the brightness of a galaxy in the observed-frame $K_{s}$ band and the massive RSGs observed in lower
redshift clusters. 

To understand this connection it is necessary to appeal to our SED fitting process. Plotted in Figure \ref{fig:restframeCMDnCSMD} is the rest-frame $M_{NUV}-M_{r^{\prime}}$ CMD and
color$-$stellar mass (CSMD) for both the Cl J0227-0421 spectral members and the field galaxy sample. Overplotted here and in Figure \ref{fig:obsCMD} are colors and magnitudes derived
from BC03 stellar synthesis models generated by EZGal\footnote{http://www.baryons.org/ezgal/model}. These models were normalized to a lower redshift ($z\sim0.6$) $L^{\ast}$ cluster 
galaxy in the observed-frame $F814W$ band (De Propris et al. 2013) and generated for a variety of different formation epochs and at a variety of different metallicities. As a rough check, we note that 
the galaxy properties generated by these models show broad agreement in the observed-frame K band with $L^{\ast}$ galaxies at similar redshifts ($z\sim3-4$) observed in 
photometric surveys and in simulations (e.g., Cirasuolo et al. 2010, Henriques et al. 2012, Muzzin et al.\ 2013). 

While effects of dust can be significant in both the CMDs and the CSMD 
(see, e.g., Lemaux et al.\ 2013) and, indeed, have been invoked as the primary culprit
for the origins of incipient protocluster red sequences observed at high redshift (Overzier et al.\ 2009), the comparisons that will be made here are differential. 
As such, it is only necessary for our purposes that the dust properties of the protocluster members not differ, on average, from those in the field at the same 
redshifts. In an attempt to ensure that this assumption holds, a stellar mass cut of $\mathcal{M}_{s}>10^9$ $M_{\odot}$ is imposed on all galaxies plotted in
Figure \ref{fig:restframeCMDnCSMD}, which, as mentioned in Appendix A, is the rough limit to which the VUDS spectroscopic sample should be representative at these
redshifts. This cut was made for two reasons, both of which are predicated on the possibility of a relationship between stellar mass and SFR suggested by a variety of observations
at a variety of epochs (e.g., Brinchmann et al.\ 2004; Daddi et al.\ 2007; Elbaz et al.\ 2007; Noeske et al.\ 2007; Santini et al.\ 2009; Gonz$\rm{\acute{a}}$lez et al.\ 2011;
Koyama et al.\ 2013). Since there is a known relationship between the SFR and the dust content of a galaxy, making this cut ensures that, to the best of our ability, the
two samples have the same average dust content. The second reason is to ensure fair comparisons between the SFRs of the protocluster members and the field population discussed
later in this section.

The large span in the protocluster galaxy properties observed previously in Figure \ref{fig:obsCMD} now appears as a bimodality in both panels of Figure \ref{fig:restframeCMDnCSMD}. There is a clear 
population of lower mass, lower luminosity, blue galaxies in the protocluster bounds that share these properties with the bulk of the field sample. There remain, however, several
spectral members that are more luminous, more massive, and redder than the overall population. These three galaxies, which 
we refer to hereafter as ``proto-RSGs'', have colors that are consistent with the last major 
star-formation event ending $\sim300$ Myr in the past, i.e., $z_{f}\sim3.75$. This formation epoch is consistent with the formation epoch derived for massive RSGs in lower 
redshift ($z\sim1-2$) clusters (e.g., Rettura et al.\ 2010; Raichoor et al.\ 2011; Hilton et al.\ 2012; Lemaux et al.\ 2012; Zeimann et al.\ 2012). The amount of stellar mass 
already in place for these galaxies at this redshift is immense considering the short period of time that they have had to form their stellar content. These 
stellar masses approach those of $z\sim1$ BCGs (e.g., Stott et al.\ 2010; Lidman et al. 2012; Ascaso et al.\ 2014), astounding considering these galaxies have $\sim$4 Gyr to evolve between 
the redshift of Cl J0227-0421 and that of $z\sim1$ cluster samples. The presence of these massive galaxies within Cl J0227-0421 appears, at least at the surface, to be consistent with 
the results of Diener et al.\ (2013), in which an excess of massive galaxies was found within protocluster candidates in the COSMOS field at slightly lower redshifts ($1.8<z<3.0$) 
and in slightly less massive structures ($\sigma_{v}=30-550$ km s$^{-1}$). This result is also broadly consistent with overdensities of bright or massive red galaxies among the 
populations of $z\sim2$ protoclusters observed by Kodama et al.\ (2007) and Zirm et al.\ (2008). Interestingly, however, the overdensity of such galaxies appears to diminish in 
protoclusters at redshifts most comparable to Cl J0227-0421 in the former study ($z\sim3$), a trend that does not seem to extend to the galaxy population of Cl J0227-0421.

\begin{figure}
\plottwoveryspecial{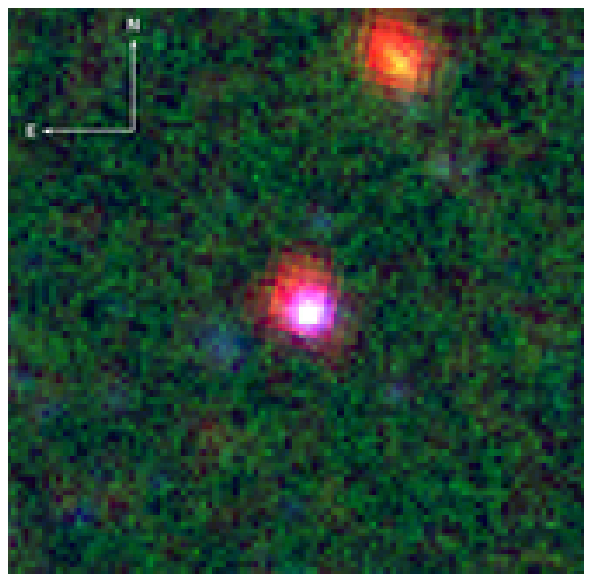}{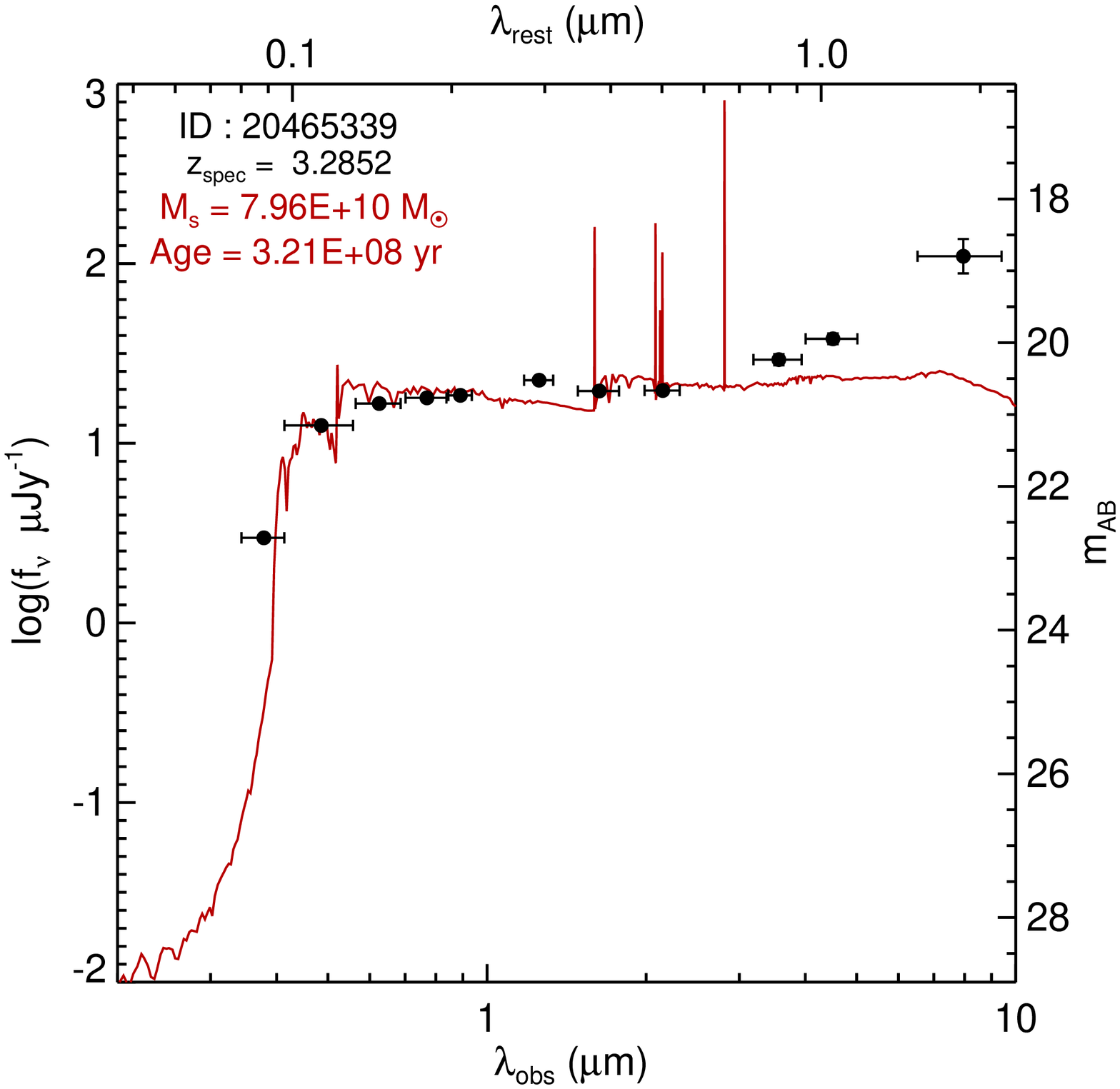}
\caption{\emph{Left:} Postage stamp of the X-ray and type-1 AGN host proto-BCG of Cl J0227-0421. Blue, green, and red channels are assigned to 
the $i^{\prime} K_{s}[4.5]$ bands, respectively, and logarithmic scaling is used for the brightness. Image dimensions are 25$\arcsec$ ($\sim200$ kpc at $z\sim3.3$) 
on each side. The bright object to the north-northwest of the proto-BCG was not targeted by spectroscopy, but has a well-fit photometric redshift of $z_{phot}=1.21\pm0.03$.
\emph{Right:} Observed-frame optical/NIR spectral energy distribution (SED) of the proto-BCG with the best-fit galaxy template overplotted in red. The proto-BCG has a power-law continuum 
both in the UV and in the IR, suggesting the physical parameter estimates coming from the SED fitting process are contaminated by the presence of the AGN. The best-fit stellar mass
and luminosity-weighted stellar age are shown in the upper left corner of the plot.}
\label{fig:BCG}
\end{figure}

\begin{figure}
\plottwoveryspecial{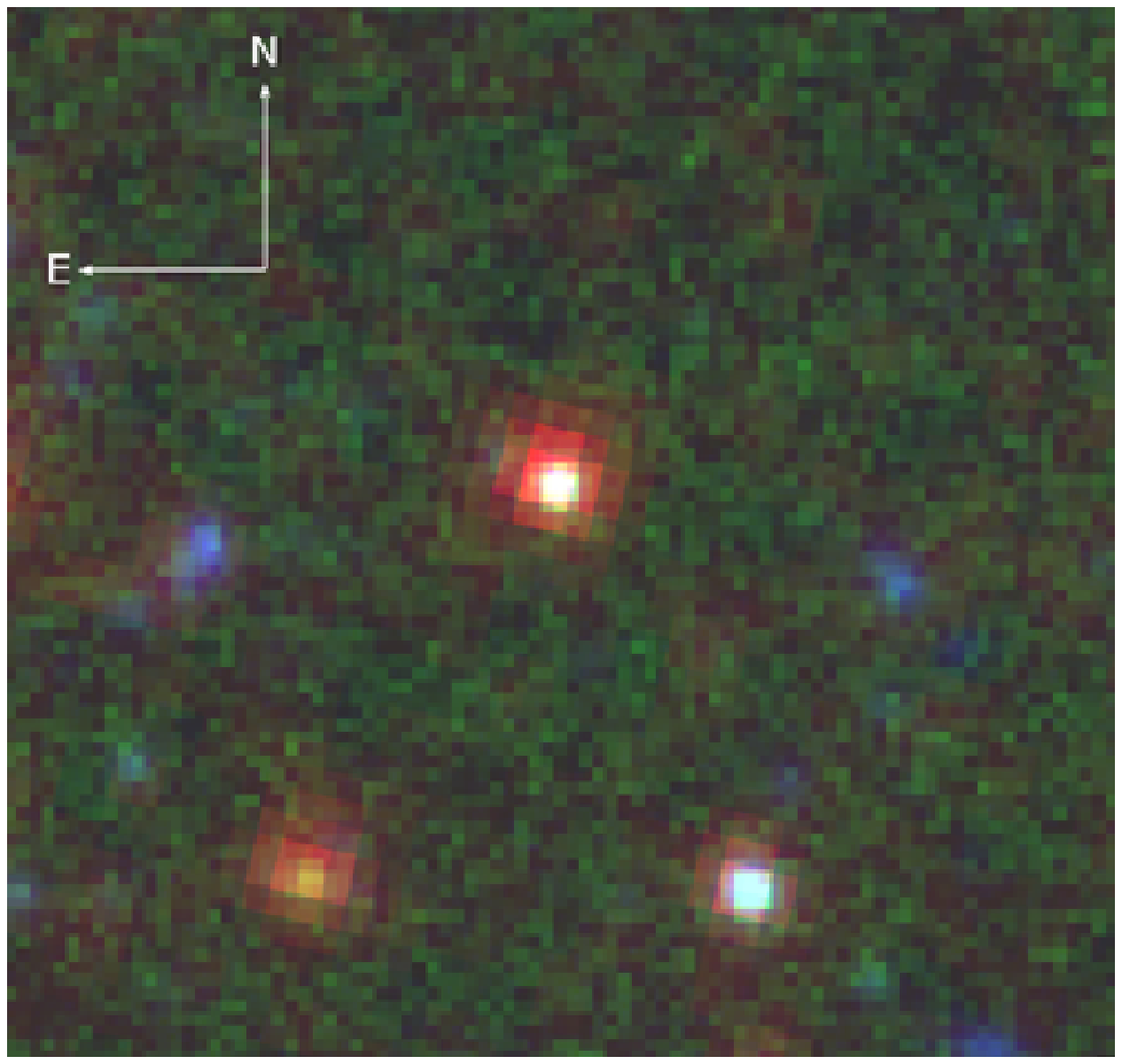}{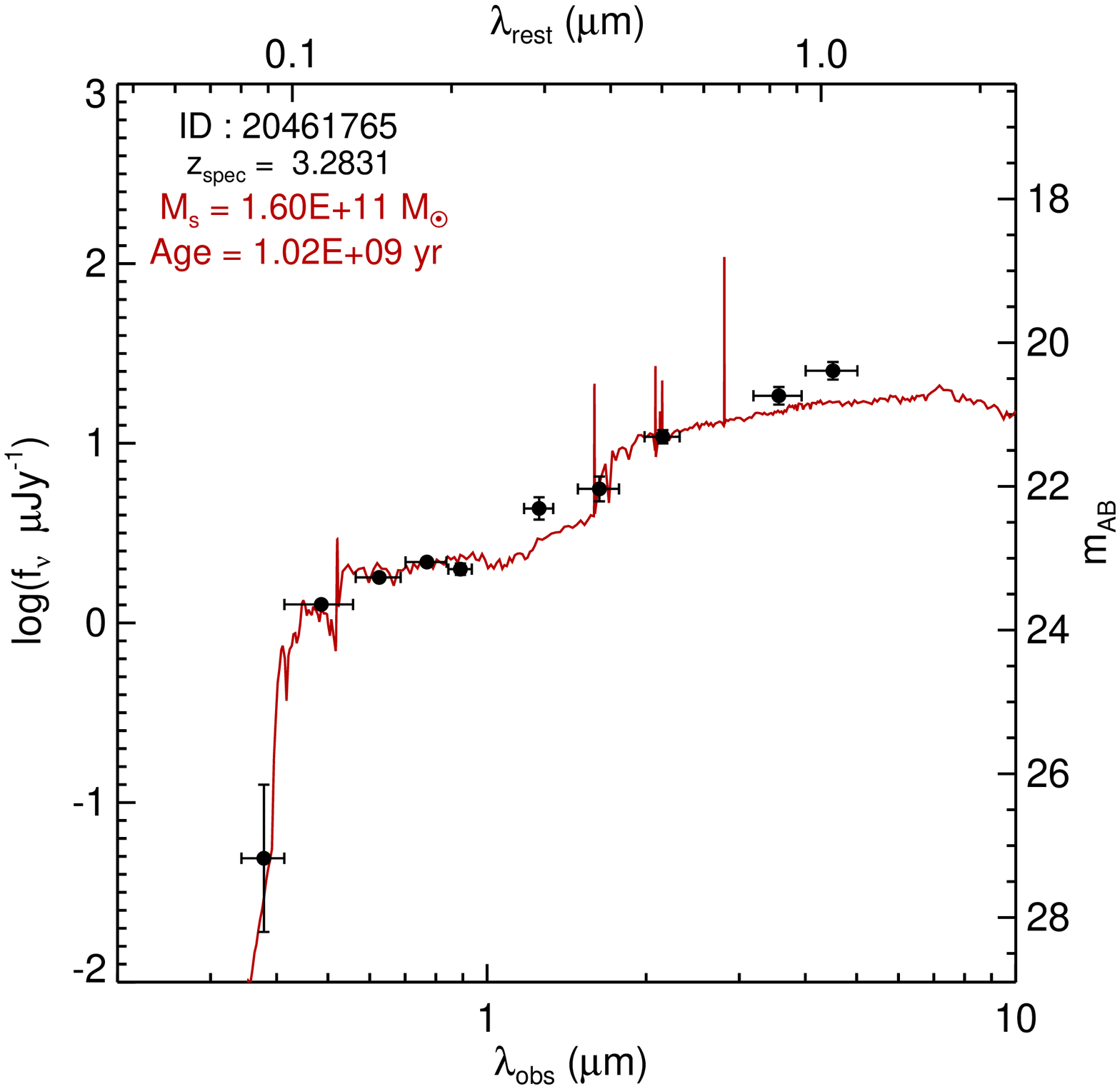}
\caption{\emph{Left:} Postage stamp of the most massive galaxy in Cl J0227-0421 generated in an identical manner to that of Figure \ref{fig:BCG}. \emph{Right:} Observed-frame 
optical/NIR SED of this galaxy plotted against the best-fit galaxy template. As before, the best-fit stellar mass and luminosity-weighted stellar age are shown in the upper left
corner of the plot. While this galaxy is also host to a type-1 AGN, the properties of its SED are much different than that of the proto-BCG and the SED is generally 
well fit by the galaxy template. There is some discrepancy in the IR portion of the SED, but an identical stellar mass is recovered (within the random errors) if IRAC bands 
are excluded from the fit.}
\label{fig:MMCG}
\end{figure}

\begin{figure}
\plottwoveryspecial{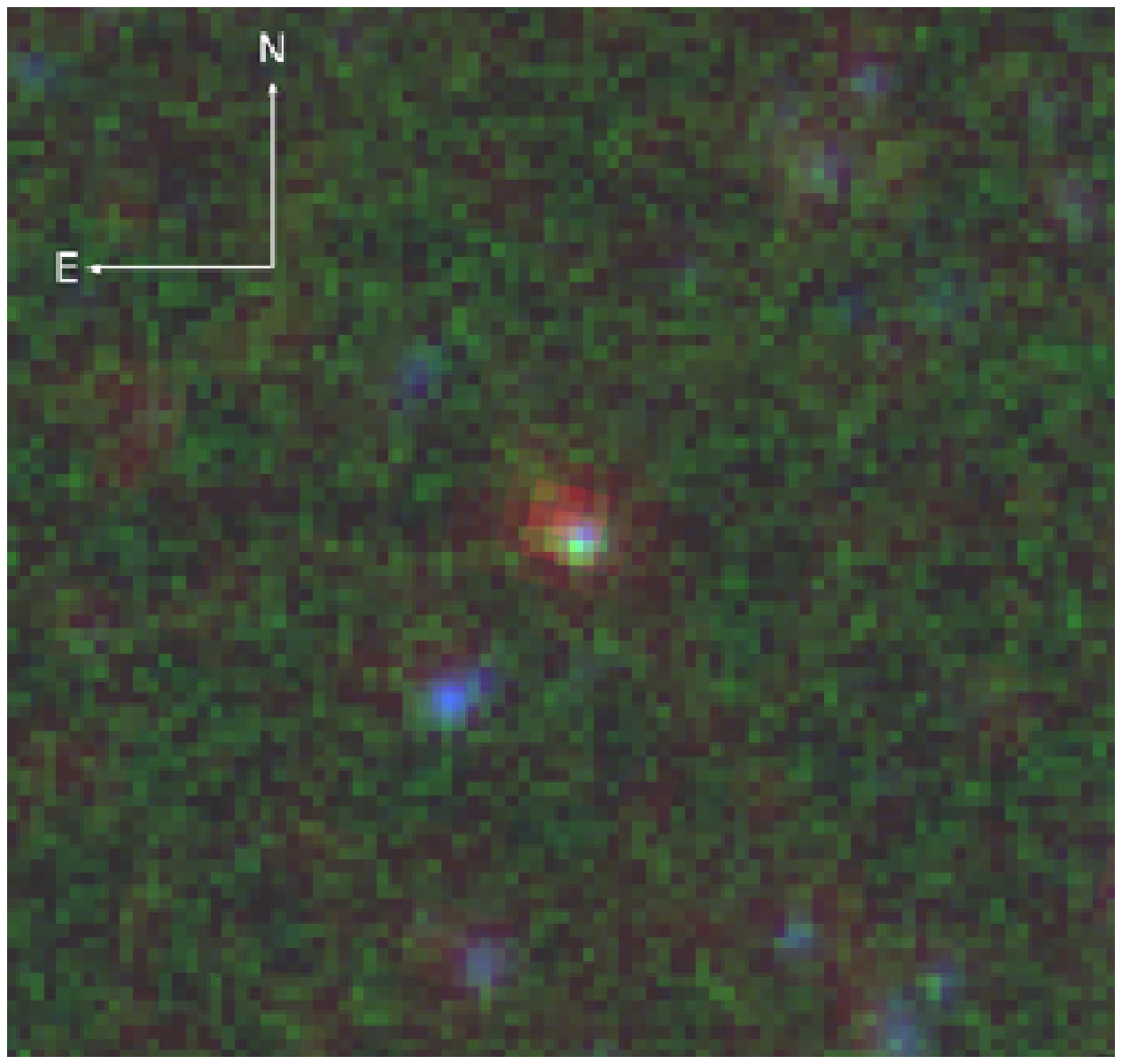}{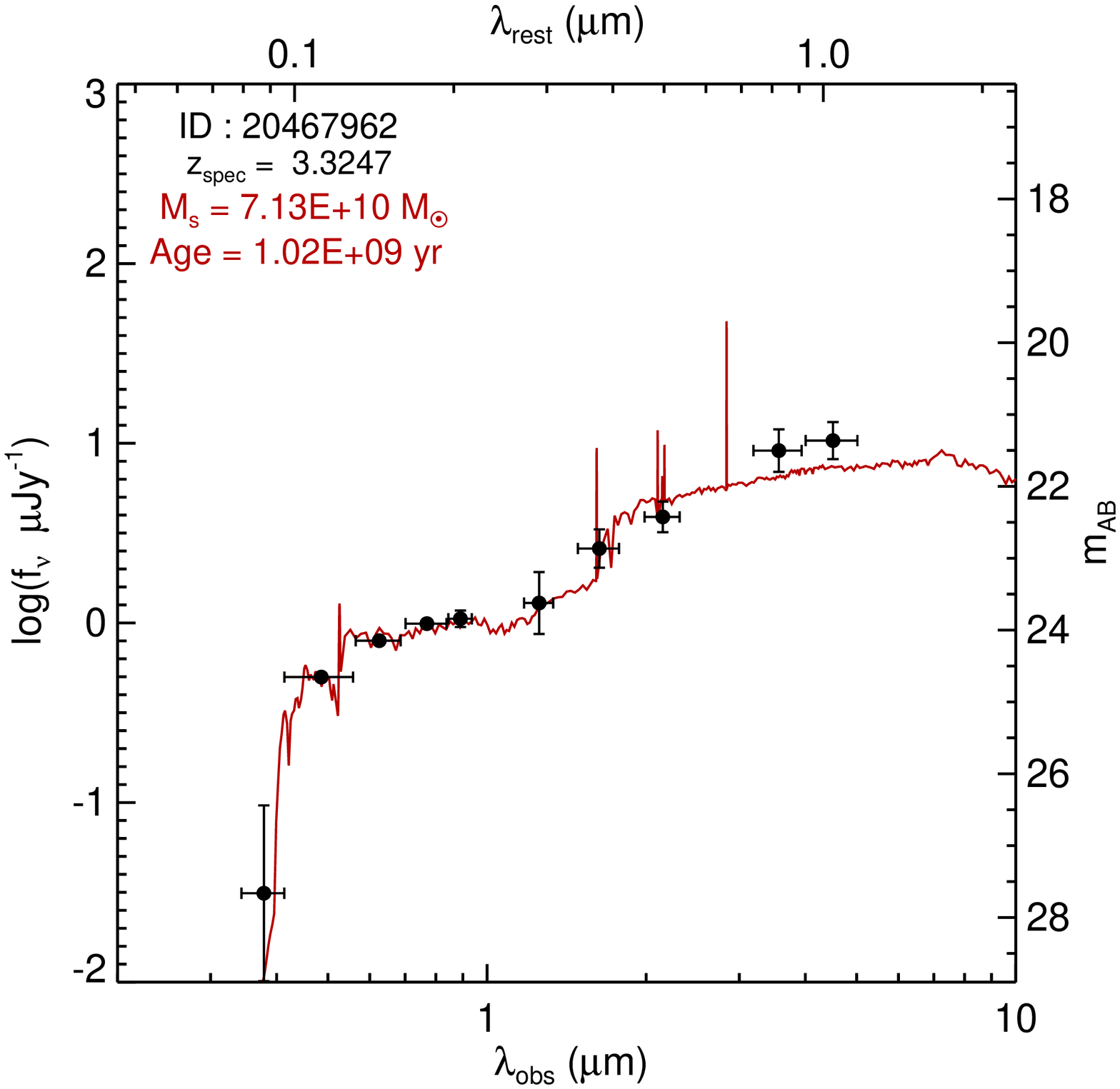}
\caption{\emph{Left:} Postage stamp of the third brightest and fourth most massive galaxy in Cl J0227-0421 generated in an identical manner to that of Figure \ref{fig:BCG}.
\emph{Right:} Observed-frame optical/NIR SED of this galaxy plotted against the best-fit galaxy template. Best-fit stellar mass and luminosity-weighted stellar age are shown in 
the upper left corner of the plot. While this galaxy is also host to a type-1 AGN, the same test was performed on this galaxy as was performed on the galaxy shown in Figure 
\ref{fig:MMCG} and negligible differences in the derived stellar mass were found.} 

\label{fig:3rdMMCG}
\end{figure}

\begin{figure}
\plottwoveryspecial{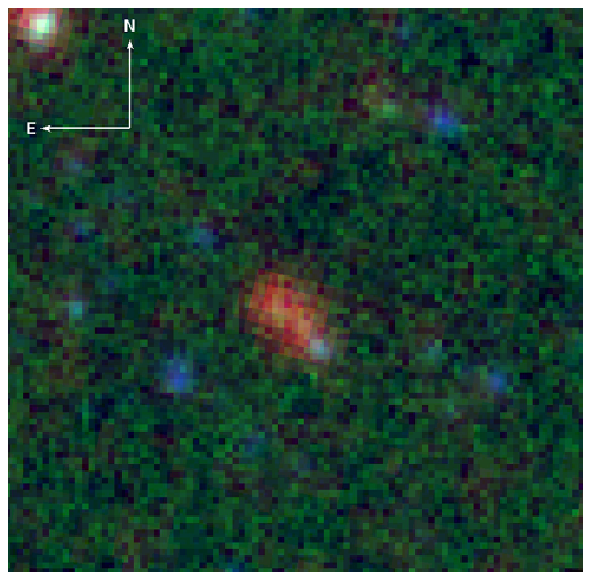}{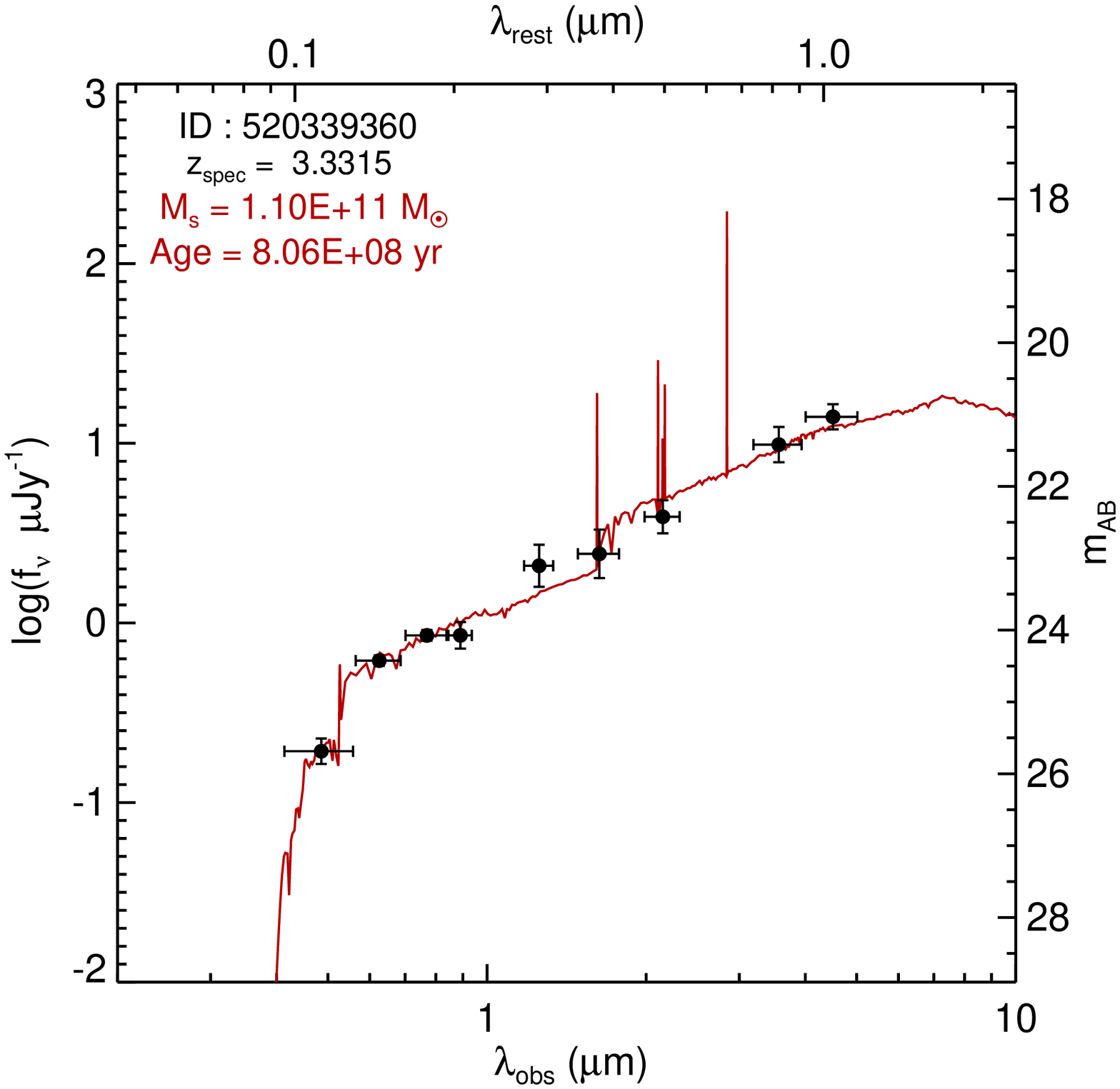}
\caption{\emph{Left:} Postage stamp of the fourth brightest and second most massive galaxy in Cl J0227-0421 generated in an identical manner to Figure \ref{fig:BCG}.
\emph{Right:} Observed-frame optical/NIR SED of this galaxy plotted against the best-fit galaxy template. Best-fit stellar mass and luminosity-weighted stellar age are shown in
the upper left corner of the plot. This galaxy is not host to a type-1 AGN. Though the best-fit galaxy template appears significantly different from the other two massive, redder protocluster 
galaxies plotted in Figures \ref{fig:MMCG} and \ref{fig:3rdMMCG}, the dominant stellar population in this galaxy appears relatively old.}
\label{fig:2ndMMCG}
\end{figure}

There are, however, two considerations. The first is a practical consideration regarding the SED fitting process. Of the four most massive galaxies
in Cl J0227-0421, three of them, including, as mentioned earlier, the proto-BCG, host broadline type-1 AGN (see Figure \ref{fig:specmosaic1}). 
There is then some concern that these three galaxies have their stellar masses, luminosities, 
and colors contaminated by the presence of broadband emission of the AGN. For the proto-BCG the estimate of the physical parameters is quite clearly contaminated by the presence
of the AGN because the SED exhibits a power-law continuum both in the 
ultraviolet and the observed-frame NIR. (This galaxy is also detected in the reddest of the IRAC bands in the SWIRE imaging.) Thus, the luminosity and color of its stellar content, as well
as the estimated stellar mass cannot be considered to be reliable. However, in the remaining two cases, the galaxy template is well-matched to the observed SED of the AGN hosts. 
These fits are shown in Figures \ref{fig:MMCG} and \ref{fig:3rdMMCG}, along with $i^{\prime} K_{s}[4.5]$ RGB postage stamp of each AGN host. Also plotted in Figure \ref{fig:2ndMMCG} 
are the SED fit and postage stamp of the one proto-RSG not host to a type-1 AGN. 

For the two proto-RSGs host to a type-1 AGN, only a few minor discrepancies with the 
galaxy template are apparent, most notably in the IRAC bands. The concern here is that a dust-obscured AGN (AGNs that are generally not typically associated with type-1 
broadline AGN) would be contributing appreciably to the rest-frame NIR luminosity, thus leading to an erroneously high stellar mass measurement. The contamination from such 
an AGN, however, decreases precipitously below rest-frame wavelengths of $\lambda_{rest}<1-2\mu m$ (e.g., Sajina et al. 2012). 

To test for the presence of contamination, we  
measured the stellar mass eliminating bands near this limit (i.e., the two IRAC bands) and measured the stellar mass again and found negligible differences ($\sim0.1$ dex) 
with respect to the fiducial fit. While the lack of more information in the NIR prevents further testing, it is sufficient to say that there is no reason to believe with
the current data that the properties of these galaxies presented in this section are affected by the presence of their AGN. In future work, hybrid galaxy/AGN templates will be
used to investigate the effect of modeling on the derived physical parameters of such galaxies further (as in, e.g., Salvato et al.\ 2009). Though there is much uncertainty in this process,
we note that the fits of the two non-proto-BCG type-1 AGN hosts yield luminosity-weighted stellar ages that are $\sim1$ Gyr in both cases, which bolsters the conclusions of 
the analysis in the CMD and CSMD that these galaxies contain a dominant older stellar population. It is also possible to use the presence of AGN 
in these galaxies to our advantage. That two-thirds (i.e., two out of three) of the proto-RSGs that have ceased to form stars several 100 Myr to a Gyr ago are host to 
a powerful AGN suggests a possible connection between AGN and quenching at these redshifts. 

Such an excess is also seen among the field population, though at a lower level: 
16.1\% of proto-RSGs in the field host a type-1 AGN, as compared to only 2.3\% for the entire mass-limited field sample. The connection between AGN and the formation of the
emergent red sequence is strengthened by the observation that AGN also appear to be more prevalent, in general, in high-redshift protocluster environments than in coeval field 
environments (see the review in Martini et al.\ 2013 and references therein). This is a trend that appears to hold for the galaxy population of Cl J0227-0421, at least among 
the proto-RSGs.

The second consideration is more expansive. A close inspection of both panels of Figure \ref{fig:restframeCMDnCSMD} reveals a large population of field galaxies with similar
luminosities, colors, and stellar masses to the massive proto-RSGs in Cl J0227-0421. The question is then does the presence of the higher density environment 
significantly enhance the number of proto-RSGs, or are the proto-RSGs observed in Cl J0227-0421 simply a sampling of similar field galaxies at this redshift? To answer this 
question, we calculated the volume probed by the spectral surveys (excluding ORELSE) at these redshifts and calculated the density of massive ($\log(\mathcal{M}_{s})>10.8$)
and red ($M_{NUV}-M_{r^{\prime}}$>1.4) proto-RSGs in both Cl J0227-0421 and the field. The $M_{NUV}-M_{r^{\prime}}$ color limit was adopted to 
roughly differentiate those galaxies whose dominant stellar population has an age in excess of $\sim200$ Myr for all models plotted in Figure \ref{fig:restframeCMDnCSMD} to those with dominate younger 
generations of stars. This criterion is not, however, sufficient to designate such populations as ``passive'' (see, e.g., Ilbert et al.\ 2013, Arnouts et al.\ 2013), but is only sufficient 
to ensure that the last major star-formation event of these galaxies was several 100 Myr in the past. 

We stress here that the quantities relating to this population that follow
were derived extremely roughly, and it will be necessary to refine this estimate once the full spectroscopic selection function and composite survey geometry has been quantified. Again, however, 
we are saved here by a relative comparison between the protocluster members and the field, which were selected in the same way from the same surveys. The space density of proto-RSGs in the field was found to be 
$\rho_{pRSG,\, field}=3.1\pm0.6\times10^{-4}$ $h_{70}^{-3}$ Mpc$^{-3}$, where errors were calculated from Poisson statistics. 

Despite the crudeness of the calculation, or, rather,
given its crudeness, this number is remarkably similar to the space density found for massive galaxies at similar redshifts in wide-field photometric surveys (Ilbert et al.\ 2013; 
Muzzin et al.\ 2013). In contrast, the space density of proto-RSGs in the bounds of Cl J0227-0421 is much higher: $\rho_{pRSG,\, pcl}=7.9\pm4.5\times10^{-3}$ 
$h_{70}^{-3}$ Mpc$^{-3}$. This difference is in large part due the volumes used to define the two samples, the volume spanned by the field sample being $\sim250$ times larger
than that of the spectral members of Cl J0227-0421. The quotient of the two densities yields a proto-RSG density contrast of $\delta_{pRSG}=25.1\pm15.2$. In other words, sampling a volume equal to that 
of the protocluster in a random part of the field would typically result in 0.12 proto-RSGs, or roughly one proto-RSG in every eight protocluster-sized volumes. Instead, three
such galaxies are found within the protocluster volume. This is perhaps unfair, however, as there are also more \emph{total} galaxies within the protocluster bounds as attested to 
by the high value of $\delta_{gal}$ found for Cl J0227-0421. The proto-RSG overdensity holds, however, if we instead consider the fraction of such galaxies in the two environments, since 
their fraction among the mass-limited protocluster member sample is nearly triple that of the field (20\% and 6.9\%, respectively). While the uncertainties on these quantities are 
extremely large owing to the small number of spectral members, the observed overdensity of proto-RSGs within the confines of Cl J0227-0421 appears to be real. This line of thought 
will continue to be expanded with the full VUDS sample.

\subsection{General properties of the Cl J0227-0421 galaxy population}
\label{generalprop}

We now move from considering the brightest and most massive galaxies in the protocluster to the spectral member population as a whole. At this point in the analysis of VUDS data,
few metrics exist, and even fewer have been extensively tested, which would allow us to attempt to separate the general member population from the field sample as a whole. 
One of these metrics, stellar mass, has already been discussed. However, when considering the spectral member sample as a whole, the distribution of stellar masses of the bulk
of the member population do not appear, by eye, to differ appreciably from the field population as a whole. This statement is quantified by a Kolmogorov-Smirnov (KS) test, which 
reveals no significant difference between the stellar mass distributions of the two samples. The second metric is the SFR of those galaxies that comprise the two populations 
as derived from the SED fitting. As discussed in \S\ref{SED}, the SFRs of a given galaxy are subject to a high level of uncertainty coming from the choice of models. 
Additionally, while it has been shown that, at lower redshift ($z\sim1$) with a great deal of care, SFRs derived from SED fitting can exhibit enough 
precision with respect to more traditional star-formation proxies (Mostek et al.\ 2012), SED-fit SFRs have not been tested enough at higher redshift. 
Therefore, we rely only on differential ensemble properties when attempting to gain any insight from these SFRs. The distribution of 
SFRs between the two samples formally shows no significant difference when compared by a KS test. However, as illustrated throughout the section with regard to stellar mass, a null result
with a KS test does not necessarily preclude the possibility that the properties of the two populations do not differ in some way. 

With a small number of galaxies in the spectral member sample, we focus instead
on the median SFRs of the two samples because this quantity does not suffer the same limitations as a KS test or a sample mean for small samples with respect to the extreme ends of the distribution. 
Significance intervals for the median SFR of each sample were calculated by bootstrapping. Shown in the lefthand panel of Figure \ref{fig:restframeCMDnCSMD} are the median SFRs for the field and 
protocluster member samples. Again, however, the small number of galaxies in the latter sample confounds interpretation. The median level of star formation activity among the protocluster galaxies appears 
suppressed relative to the field by roughly a factor of two (i.e., $SFR_{med}\sim30$ vs. $\sim55$ $M_{\odot}$ yr$^{-1}$, respectively)\footnote{At this stage 
of the analysis, the tentative \emph{Herschel}/SPIRE detection of the proto-BCG is not incorporated due to its ambiguous nature and the lack of matched 
\emph{Herschel}/SPIRE data for the full field sample.}, but the large errors mean that this result is significant only at the $\sim1\sigma$ level. Moving beyond 
the formal statistical errors, the difference in SFRs between the two samples remains significant at $\ge1\sigma$ (with the 
same directionality) if other SFHs are employed or if the mean is adopted rather than the median. Still, however tempting, we cannot make any definite claims with the current data and 
wait instead to incorporate the large sample of galaxies in overdense environments in the early universe afforded by the full VUDS sample to see whether this trend gains significance.

\begin{figure*}
\epsscale{0.7}
\plottwo{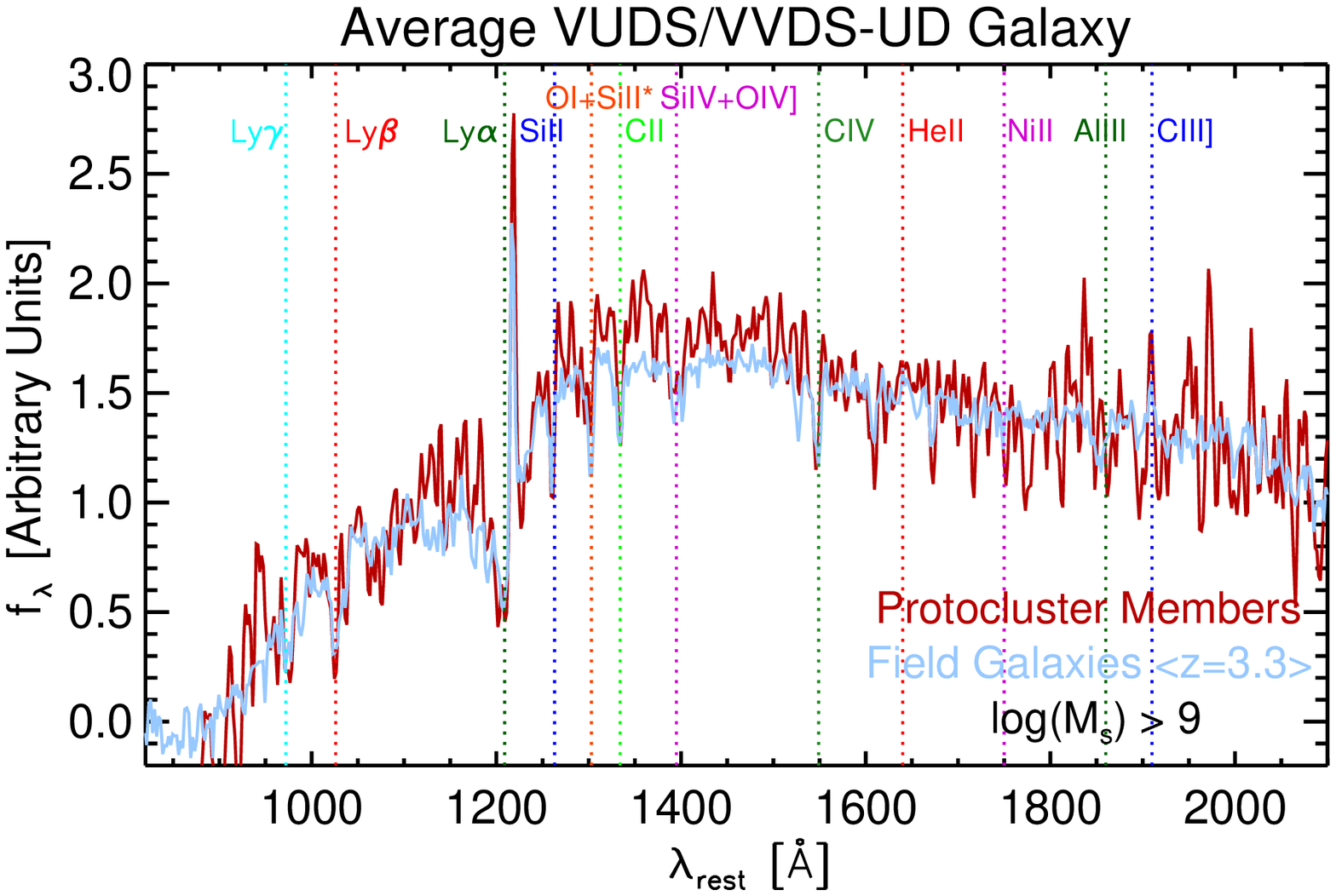}{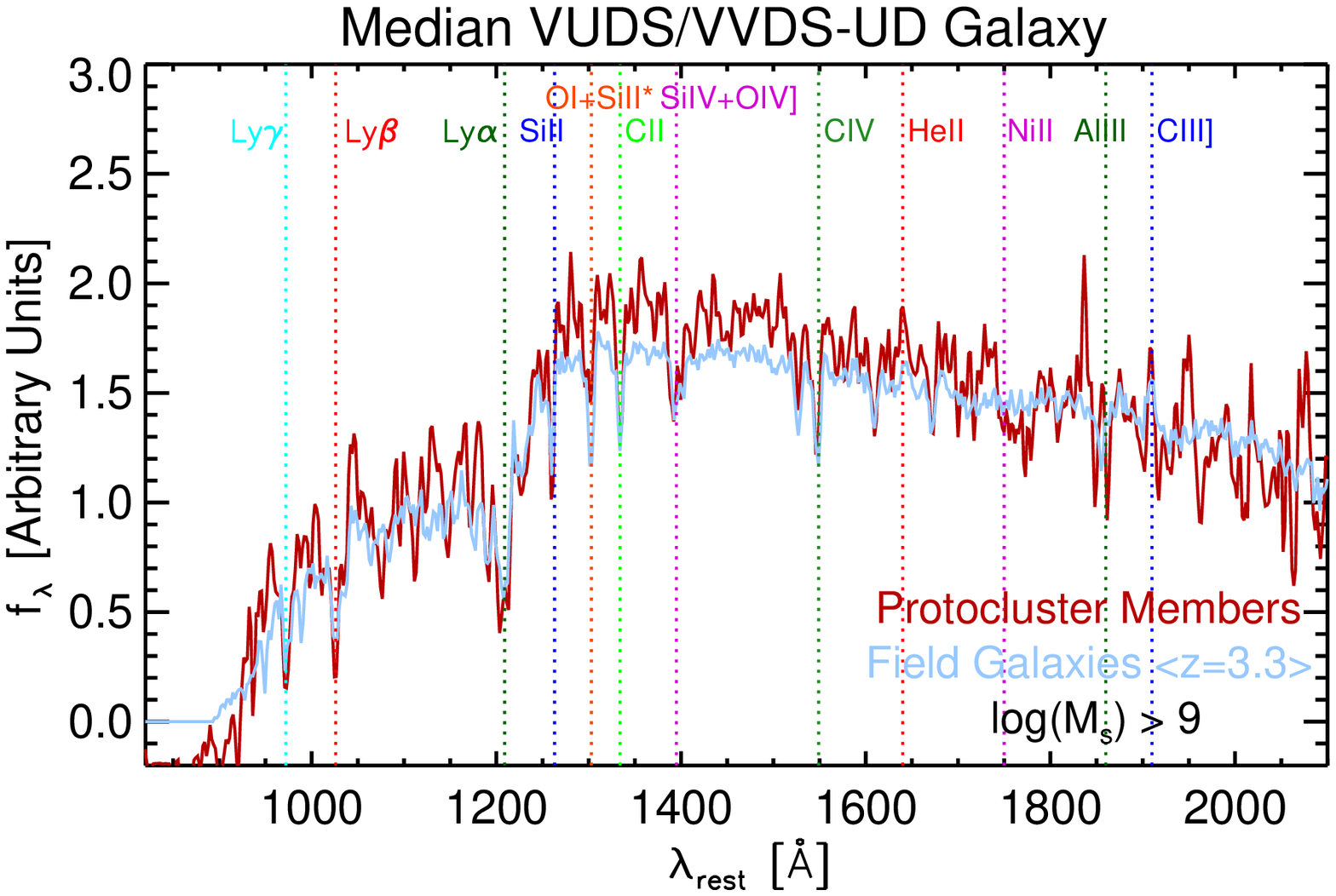}
\caption{\emph{Left:} Rest-frame mean ``coadded'' VUDS spectra of all spectral members of Cl J0227-0421 with $\log(\mathcal{M}_s)>9$ plotted 
against that of the field galaxy sample. All type-1 AGN hosts are removed from both samples. Important spectral features are noted. The two 
spectra exhibit remarkably similar properties. Subtle differences do exist; the average protocluster galaxy appears to have slightly stronger emission features (Ly$\alpha$,
HeII, and CIII]) and a steeper rest-frame NUV continuum slope. \emph{Right:} Median ``coadded'' VUDS spectra of the same two samples shown in the left panel. The major difference between the median spectra 
and the mean spectra plotted in the left panel is the enormous decrease in the strength of the Ly$\alpha$ line, which is a reflection of the fact that most galaxies
do not emit Ly$\alpha$ at these redshifts. In the median spectrum, the Ly$\alpha$ emission is roughly equivalent among the field galaxies and the protocluster members,
owing to the similar fractional contribution of LAEs to both samples (see text). The differences in the median and mean spectra suggest that, though the galaxies 
in the protocluster emit Ly$\alpha$ roughly as frequently as the general field population, those galaxies which do emit Ly$\alpha$ in the protocluster are doing so 
more profusely.} 
\label{fig:coaddmosaic}
\end{figure*}

The final tool that we have at our disposal is the rest-frame NUV spectra of both the members of Cl J0227-0421 and field galaxies at the same epoch. A preliminary and simple method of implementing
this tool is to count the number of galaxies exhibiting emission lines (in this case Ly$\alpha$) vs. those that do not. Such an exercise can perhaps give information about the constituent 
galaxy populations of the two samples since LAEs have been shown to differ appreciably in their SFRs, luminosities, and in the ages of their dominant stellar population with respect 
to Lyman break galaxies that do not emit in Ly$\alpha$ (e.g., Shapley et al.\ 2003; Lai et al.\ 2007; Pentericci et al. 2007, 2009; Kornei et al. 2010). However, no clear excess in LAE fraction 
is observed for the spectral members of Cl J0227-0421. Among the members that have spectra with sufficient wavelength coverage, $33\pm15$\% of the members exhibited 
rest-frame Ly$\alpha$ equivalent widths (EWs) in excess of 25\AA, a fraction consistent with the average fraction among a volume-limited sample of VUDS galaxies at these redshifts. 
For further details on the measurements of Ly$\alpha$ EWs and the average LAE fraction, as measured from VUDS, over a large redshift baseline, see Cassata et al.\ (2014). 

A slightly more complicated method of implementing this tool is through the combination of spectra to create a high signal-to-noise ratio (S/N) spectrum that in some way 
represents the average properties of the constituent galaxies, a process we refer to hereafter as ``coadding'' (with the resulting product referred to as a ``coadded spectrum'').
After removing all type-1 AGN hosts, the galaxies comprising the two samples were coadded separately, loosely following the methodology outlined in Lemaux et al.\ (2013).
Here, however, we chose not to use weighting based on the formal uncertainty spectra generated for VUDS by the VIMOS pipeline (see Le Le F{\`e}vre et al.\ 2014 and references therein for details), 
because the properties of these spectra have not been extensively tested to date. The effect of not weighting each input flux density by its uncertainty (or variance) 
is to appreciably add noise to the continua of the coadded spectra in the regions of under- or oversubtracted airglow lines. The added noise in the resulting coadded spectra in turn leads to 
increased uncertainty in physical parameters derived from the spectra. This problem is compounded for small samples and samples that span limited redshift intervals. Though we attempted 
to mitigate this effect in some way by also generating coadded spectra based on the median flux density of input spectra at each pixel, we restrict ourselves here 
to a broad qualitative comparison of the coadded spectra, turning only very briefly to a quantitative comparison. 

Plotted in Figure \ref{fig:coaddmosaic} is the mean (hereafter ``average'') and median 
coadded spectra of members of Cl J0227-0421 not host to a 
type-1 AGN against the backdrop of those of the field sample. Over the region where the airglow contamination is minimal ($\lambda_{rest}<1800$ \AA), both sets of coadded spectra (left panel of 
Figure \ref{fig:coaddmosaic}) exhibit remarkably similar properties. Both the Ly$\alpha$ absorption trough, and many ISM absorption features are of comparable relative strength 
in the two sets of coadded spectra. One broad difference appears to be the slope of the NUV continuum, which appears steeper in the both the average protocluster galaxy and in the median coadded spectrum
of protocluster members relative to the field. While highly subject to both the amount, geometry, and composition of dust in a galaxy, the IMF, and stellar metallicity (e.g., Castellano et al.\ 2012; 
Wilkens et al.\ 2012), this quantity, 
typically referred to as the $\beta-$slope, can be linked to the SFRs, stellar masses, and mean luminosity-weighted stellar ages of galaxies. 

While several recent measurements of this 
quantity in high-redshift galaxies have been made with photometry (e.g., Bouwens et al.\ 2012; Dunlop et al.\ 2012; Finkelstein et al.\ 2012; Jiang et al.\ 2013; Hathi et al.\ 2013; Castellano et al.\ 2014),
similar measurements with spectroscopy have not, to date, been attempted for a large population of such galaxies. The relationship of the $\beta-$slope with galaxy properties is an area 
of active investigation with VUDS (Hathi et al.\ 2014), and these results will be used in the future, along with the full VUDS sample to interpret and contextualize differences 
in the $\beta-$slope as a function of environment. Another possible difference between the average spectral properties of the two populations lies in the strength of the 
emission features, with protocluster galaxies having, on average, slightly stronger Ly$\alpha$, HeII, and CIII] emission. Since, as noted above, the two populations have comparable 
fractions of LAEs, this possibly suggests a connection between environment and Ly$\alpha$ escape fraction, as well as possible AGN activity. However, with the limited sample presented here,
it is not possible to make definitive claims, and we instead wait for the inclusion of the full sample of VUDS overdensities to explore this thought rigorously.

In an attempt to make a slightly more quantitative comparison, both coadded spectra were fit using GOSSIP to BC03 synthetic spectral models that spanned an 
identical parameter space to those used for the SED fitting process. Because the method of coadding spectra destroys absolute flux calibration through normalization, without 
a more complicated implementation of similarly averaged photometry for each sample, only 
relative quantities, i.e., stellar age, metallicity, and extinction, can be used from these fits. Though large uncertainties exist in the fitting process, the best-fit models 
to the two average coadded spectra yield parameters that also show a high degree of similarity. The average field and protocluster galaxies have identical metallicities and 
luminosity-weighted stellar ages and stellar extinctions that differ only by one resolution element (300 Myr and 400 Myr and $E(B-V)=0.3$ \& 0.2, respectively). These parameters 
do not change appreciably when the median coadded spectra are fit instead (i.e., 300 Myr and 300 Myr and $E(B-V)=0.3$ \& 0.2, respectively).

Within the limitations of the analysis presented in this section on average properties, the galaxies comprising the Cl J0227-0421 protocluster appear to be 
broadly similar to those in the field at similar redshifts. While there was marginal evidence of the suppression of star-formation activity within the protocluster 
boundaries, as well as differences in the average $\beta-$slope and average emission line strengths, all other ensemble quantities showed broad concordance between the two samples. There was a significant
difference, however, between the two samples when a specific population was isolated, with a marked excess of massive, red galaxies, observed in the protocluster 
environment. Among these galaxies, we also observed tentative evidence of increased AGN activity, including in the extremely bright and blue proto-BCG. With the 
large amount of total mass already assembled in Cl J0227-0421 at this redshift (see \S\ref{halomass}), these results hint at a picture where we are witnessing 
the birth of quenching within the protocluster environment, with massive galaxies beginning to transition to the red sequence setting the stage for future 
environmental influences on the bluer, lower mass member galaxies. While this picture will continue to be focused and contextualized with the full sample of 
VUDS protoclusters, this first look into the heart of an emerging massive cluster at high redshift has provided several enticing clues to what  
may eventually form. 


\section{Summary and conclusions}
\label{conclusions}

In this paper we have described a systematic search for overdensities at high redshift ($z>2$) in the CFHTLS-D1 field using newly obtained VUDS spectroscopic data in conjunction with
the wealth of other imaging and spectroscopic data available for this field. We then described the discovery and characterization of the most significant of these overdensities,
the Cl J0227-0421 protocluster at $z\sim3.3$. Here we briefly outline the main conclusions of this study.

\begin{itemize}
\renewcommand{\labelitemi}{$\bullet$}

\item With 19 confirmed spectroscopic members and six potential spectroscopic members, Cl J0227-0421 is significantly overdense relative to the field at these redshifts. 
Using a large field coeval population from VUDS and VVDS along with the 19 confirmed spectroscopic members, we estimated the significance of the spectroscopic overdensity 
of Cl J0227-0421 to be $\sigma=13.5$ (or $\delta_{gal}=10.5\pm2.8$). After accounting for spurious peaks, we found that Cl J0227-0421 is also overdense in its photometric 
redshift members, with a significance of $\sigma=8.0$.
 
\item Four different methods were used to estimate or place limits on the halo mass of Cl J0227-0421 at $z\sim3.3$ or $z=0$ (or both). These were member galaxy dynamics,
stellar-to-halo mass, X-ray hydrostatic equilibrium, and spectroscopic galaxy overdensity. Though the errors, uncertainties, and number assumptions used for each 
method were large, a consistent picture emerged in which Cl J0227-0421 has already assembled a large amount of mass in the early universe ($\mathcal{M}_{z\sim3.3}\sim3\times10^{14}$ 
$M_{\odot}$) and will evolve into a cluster with a halo mass rivaling or exceeding that of Coma ($\mathcal{M}_{z=0}\sim4\times10^{15}$ $M_{\odot}$).

\item The properties of the spectroscopic member galaxies of Cl J0227-0421 were investigated. In the brightest protocluster galaxy, we found evidence of a powerful active 
galactic nuclei, as well as tentative evidence of vigorous star formation activity ($\sim750$ $M_{\odot}$ yr$^{-1}$). Within the protocluster environment, a
significant excess of brighter, redder, and more massive galaxies
appeared relative to a similarly selected field population at similar redshifts. This excess was quantified both 
absolutely, $\delta_{pRSG}=25.1\pm15.2$, and relatively, with a fractional excess of such galaxies within the protocluster of around three. Based on comparisons with models, 
the last major star-formation event in these galaxies was estimated to be in excess of 300 Myr prior to $z\sim3.3$, indicating that we may be witnessing the
onset of environmentally-driven quenching processes. 

\item The remaining protocluster members had properties that were broadly similar to those of field galaxies. While we found weak evidence of suppression of the star formation 
rates among the general protocluster member population and subtle differences between the stacked spectra of the two populations, these differences were not significant 
enough to be conclusive. 
\end{itemize}

Despite the massive nature of Cl J0227-0421, the relatively small number of protocluster members statistically limited the conclusions that could be drawn. Still, the 
results of several lines of analysis presented in this paper were tantalizingly suggestive of the effect of environment at $z\sim3.3$. These lines of analysis
will be continued with the $\sim$40 overdensities found within the entire VUDS sample to search for definitive signs of environmentally-driven evolution and transformation 
in the high-redshift universe.

\begin{acknowledgements}
We thank ESO staff for their continuous support for the VUDS survey, particularly the Paranal staff conducting the observations and Marina Rejkuba and the ESO user support group in Garching.
This work is supported by funding from the European Research Council Advanced Grant ERC-2010-AdG-268107-EARLY and by INAF Grants PRIN 2010, PRIN 2012 and PICS 2013.
AC, OC, MT and VS acknowledge the grant MIUR PRIN 2010--2011.
DM gratefully acknowledges LAM hospitality during the initial phases of the project.
B.C.L. gratefully acknowledges the kindness, support, and salty muffins of Debora Pelliccia provided, even at the most unreasonable of hours, throughout the course of this work.
This work is based on data products made available at the CESAM data center, Laboratoire d'Astrophysique de Marseille.
This work partly uses observations obtained with MegaPrime/MegaCam, a joint project of CFHT and CEA/DAPNIA, at the Canada-France-Hawaii Telescope (CFHT) ,
which is operated by the National Research Council (NRC) of Canada, the Institut National des Sciences de l'Univers of the Centre National de la Recherche Scientifique 
(CNRS) of France, and the University of Hawaii. This work is based in part on data products produced at TERAPIX and the Canadian Astronomy Data Centre as part of the 
Canada-France-Hawaii Telescope Legacy Survey, a collaborative project of NRC and CNRS. A portion of the spectroscopic data presented herein were 
obtained at the W.M. Keck Observatory. We wish to thank the indigenous Hawaiian community for allowing us to be guests on their sacred mountain; we are most 
fortunate to be able to conduct observations from this site.
\end{acknowledgements}

\appendix

\section{Photometric and spectroscopic SED fitting}

For the initial spectral energy distribution (SED) fitting used to generate photometric redshifts for spectroscopically untargeted objects, only
ground-based optical and NIR photometry (i.e., CFHTLS/WIRDS) were used because the SERVS data had not been incorporated into the full photometric redshift catalog at
the time of publication. This lack of observed-frame coverage redward of $\lambda_{obs}\ga2\mu m$ is, however, of limited consequence, since 
this fitting is used here only for photometric redshifts. Redshifts derived in this manner are known, even up to the highest redshifts
of our sample, to be relatively invariant under the inclusion of \emph{Spitzer} imaging (see, e.g., Brada{\v c} et al.\ 2014; Ryan et al.\ 2014) for datasets
with broadband filters that probe both the Lyman-limit/Ly$\alpha$ break and the Balmer/4000\AA\ break. Regardless, we tested this assumption on our own data for the subset of galaxies
with secure spectroscopic redshifts at $z>2$, the redshift range of interest for this study. No statistically significant difference in the normalized absolute median
deviation, $\sigma_{\Delta z/(1+z_s)}$, (NMAD; Hoaglin et al.\ 1983), or the catastrophic outlier rate (i.e., $|z_p-z_s|/(1+z_s) > 0.15$, see Ilbert et al.\ 2013) was found 
between photometric redshifts determined with and without SERVS data included. A comparison of the photometric redshifts derived from the CFHTLS/WIRDS photometry and spectroscopic redshifts
of all objects targeted in the CFHTLS-D1 field that have a secure spectroscopic redshift yielded a catastrophic outlier rate of 9.7\%, a rate considerably higher than the 
VVDS data alone (see Lemaux et al.\ 2013). However, once catastrophic
outliers were rejected, the NMAD was $\sigma_{\Delta z/(1+z_s)}=0.030$, essentially identical to 
VVDS.  

For those objects that had been targeted with spectroscopy, the SED fitting process was performed separately on objects with and without secure spectroscopic redshifts.
For the former, spectroscopic redshifts were used as a prior to fix the redshift of the source prior to the SED-fitting, and for the latter, the redshift, as with
untargeted objects, was left unconstrained. The models used and the range of parameters were identical to those of the previous fitting and are described in Lemaux et al.\ (2013)
and references therein. The SERVS data, previously unused, was incorporated for this instance of the SED fitting as the physical parameters derived from the SED fitting are
heavily used in this study and are known to be sensitive to the inclusion of IRAC data (see discussion in Brada{\v c} et al.\ 2014). Two other changes relative to the
version of the fitting presented in Lemaux et al.\ (2013) were made at this point. The first was to use
best-fit model, i.e., the combination of template and physical parameters that minimized the $\chi^2$ with respect to photometric data, rather than the median
of the probability distribution function (PDF). This choice was made because roughly 15\% of our spectroscopic sample were significantly detected in an insufficient number of
photometric bands to satisfactorily calculate a PDF. For the 85\% of the sample for which a comparison could be made, no systematic offset was observed between the
best-fit and median stellar masses, luminosity-weighted stellar ages, and star formation rates (SFRs), with a negligibly small scatter between the two estimators of 0.04 dex for all three
parameters. The second change made in this version of the SED fitting was the use of MAG\_AUTO measurements instead of the scaled aperture magnitude 
measurements used in the Lemaux et al.\ (2013). Though the former are known to be more susceptible to blending, we preferred these measurements as they showed greater consistency with
the aperture-corrected SERVS magnitudes.

Because it has been suggested that high-redshift galaxies have star formation histories (SFHs), which deviate considerably from the simple exponentially decaying tau model (e.g., Maraston et al. 2010;
Reddy et al.\ 2012; Schaerer et al 2013; Buat et al.\ 2014; though see also, e.g., Ryan et al.\ 2014; Sklias et al.\ 2014), the effect
of changing the SFH was tested by rerunning SED fitting using Bruzual \& Charlot (2003; hereafter BC03) delayed tau models with an identical initial mass function (IMF; Chabrier 2003)
and an identical range of extinctions, taus, and metallicities. For the two physical parameters that are of paramount interest for this study, stellar mass and SFR, only a small
systematic offset of 0.05 dex between the best-fit parameters of the two different SFHs was observed, with the delayed tau model yielding slightly
higher SFRs and slightly lower stellar masses. The r.m.s. scatter between the two sets of parameters measured with the two different SFHs was also small: 0.04 dex for
both parameters. Because all of the comparisons that are made in this paper are internal, it would have no effect on our results if the SFHs of galaxies in our sample were
\emph{globally} mischaracterized to the same level. However, since we made comparisons between galaxies in different environments, it is possible
that the SFHs of galaxies depends on environment (e.g., Kauffmann et al.\ 2004), which would lead to a differential bias in the physical parameters. It is therefore comforting that, at least
for these two SFHs, the differences between the physical parameters derived for the two sets of models is negligibly small. Because of its consistency with the previous
SED fitting and to ease comparisons with the vast majority of other studies, we decided to adopt those parameters derived from the exponentially decaying tau model. With
these sets of models, the typical (random) uncertainty in the stellar mass and SFRs of galaxies in the range of interest for this study (i.e., $9<\log(\mathcal{M}_{s})<12$, 
$2.9<z_{spec}<3.7$) coming from the SED fitting process were 0.16 and 0.10 dex, respectively. 

Spectra were fit using the GOSSIP software, a package created to to fit the spectro-photometric emission of a galaxy with a set of synthetic models with a library builder that allows for
the construction of various resolution BC03 and Maraston (2005, 2011) models. For this study, we fit only exponentially decaying BC03 models with the same assumptions
and those spanning the same parameter space as those adopted for the photometric SED fitting described above. Among the many improvements that have been recently implemented on
GOSSIP, one of the most important targets for the redshift range of VUDS is related to the treatment of the intergalactic medium (IGM) extinction. While a typical assumption is to
employ the IGM model of Madau (1995), which produces, for a given redshift, a single IGM extinction curve, GOSSIP is able to choose up to five different IGM curves
along various sight lines, which provides a more realistic determination of the resultant physical parameters. Both GOSSIP and the improvements that have been
made to it for general use with the VUDS survey is explained in Thomas et al.\ (2014).

\section{Details of the halo mass estimates of Cl J0227-0421}

The initial methodology used to determine an estimate on the halo mass of Cl J0227-0421 utilized the information provided by the dynamics of the member galaxies. 
The implicit assumption in this method is that the protocluster is in a virialized state, an assumption that almost certainly does not hold at this redshift given the 
limited time member galaxies have had to interact with the potential. The high degree of skewness observed in the differential velocity distribution of the spectral members 
quoted in \S\ref{protocluster} attests to the failure of this assumption. In the case of a structure in the initial stages of its collapse, the
measured velocity dispersion will potentially decrease relative to the virial value owing to galaxies appearing compressed along the redshift
dimension (e.g., Steidel et al.\ 1998). At later stages, however, the measured velocity dispersion will be an overestimate of the virial value
as galaxies that have fallen from long distances begin to make their first passes through the protocluster core. Given the young age
of the universe at $z\sim3.3$, the former is the stronger of the two possibilities. However, with no knowledge of the true evolutionary stage
of the dynamics of the spectral members of Cl J0227-0421, we remained doubtful about this point, and simply calculated the dynamical mass with
the knowledge that this quantity can be a lower or an upper limit. The virial dynamical mass was calculated via

\begin{equation}
\mathcal{M}_{dyn, \, vir} = \frac{3\sqrt{3}\sigma_{v}^3}{11.4 G \, H(z)}
\label{eqn:Mvir}
\end{equation}

\noindent where $G$ is Newton's gravitational constant and $H(z)$ is the value of the Hubble parameter at the redshift of interest. This
formula is used directly to calculate the dynamical mass at the virial radius reported in Table \ref{tab:massprop}. 

The calculation relating the stellar mass of members of the protocluster to the total mass was done in the following manner. Loosely following
the methodology of Strazzullo et al.\ (2013), we adopt the relationship between halo mass and the stellar mass of members within $r_{200}$,
the radius at which the mean density is 200 times that of the critical density, calibrated using data from the Sloan Digital Sky Survey (SDSS) by Andreon (2012).
To determine the amount of stellar content in Cl J0227-0421, we summed the stellar masses of all spectral members with stellar masses
in excess of 10$^{9}$ $M_{\odot}$, chosen as it is roughly the turnover in number counts of all VUDS galaxies with secure spectroscopic
redshifts from $2.9<z<3.7$, the redshift bounds used to define our field sample in \S\ref{CMDnCSMD}. The large projected radius over which
we sum the stellar mass of spectral members (i.e., $R_{proj}<3$ $h_{70}^{-1}$) was motivated by the high likelihood of such galaxies becoming
as virialized members by $z\sim0$ (Chiang et al.\ 2013, Zemp 2013), the redshift at which the Andreon (2012) relation was calibrated. Since the virial
radius is typically defined to be smaller than $r_{200}$ (Biviano et al.\ 2006; Poggianti et al.\ 2009), such galaxies should be accounted for
in this relation.

A large number of the objects within the protocluster bounds are, however, not sampled spectroscopically or do not have a secure spectroscopic redshift. To
account for this lack of sampling, we calculated the probability of being a true member by comparing the spectroscopic and photometric redshifts of those objects with 
secure spectroscopic redshifts. Two probabilities were calculated, one for objects with photometric redshifts consistent with the redshift of the protocluster 
and those that were not. The correction to the composite stellar mass is then

\begin{multline}
\Sigma\mathcal{M}_{s,\, corr} (\mathcal{M}_s>10^9 \, M_{\odot}) = \Sigma\mathcal{M}_{s,\, uncorr}\\
\left(\frac{P(z_{p,\, mem}|z_{s,\, mem})N_{p,\, mem} + P(z_{p,\, nm}|z_{s,\, mem})N_{p,\, nm} + N_{s,\, mem}}{N_{s,\, mem}}\right)
\label{eqn:smcorr}
\end{multline}

\noindent where $N_{p,\, mem}$ is the number of photometric redshift members that went untargeted or that have questionable spectroscopic redshifts within
the bounds of the protocluster, and $N_{p,\, nm}$ is the equivalent quantity for photometric redshift non-members. 

The two probabilities, defined
as the likelihood of being a spectral member in the event that an object is a photometric redshift member or non-member, were determined to be
31.0\%\footnote{Though this number looks at first glance to be inconsistent with the claim in \S\ref{analysis} that
a significant overdensity of $z_{phot}$ members is likely to be real, the threshold of significance was determined empirically through a comparison of photometric
redshift source and spectral overdensities and thus accounts for this impurity.} and 0.4\%, resulting in a correction factor of 5.5. In this calculation
the assumption is made that the untargeted members have an identical stellar mass distribution to the spectral members, which is reasonable given 
that the VUDS sample should be representative of galaxy populations at these redshifts bounded by this stellar mass limit. The quantity in 
Equation \ref{eqn:smcorr} was further corrected for galaxies between $10^{8}<\mathcal{M}_s<10^9 \, M_{\odot}$ by integrating the stellar mass
functions (multiplied by $\mathcal{M}_{s}$) derived by Ilbert et al.\ (2013) for galaxies at $3<z<4$. The lower limit of this correction is set by the rough stellar mass
completeness limit of SDSS at the redshift of those clusters used for calibration (Panter et al.\ 2007). The resultant total corrected stellar mass is
3.22$^{+0.20}_{-0.32}\times10^{12}$ $h_{70}^{-1}$ $M_{\odot}$, where the errors were determined from the SED fitting process. It is interesting to note that this 
corrected total stellar mass rivals the total stellar mass contained in red-sequence galaxies in massive, $z\sim1$ clusters (e.g., Overzier et al.\ 2009; 
Rettura et al.\ 2010; Lemaux et al.\ 2012). Under the assumption that the correction made here is appropriate, and under the additional assumption that $\sim$4 Gyrs 
of evolution within a (proto) cluster environment is enough to, by and large, quench member galaxies, it appears that a large amount of the requisite stellar mass
needed to populate the red sequence of lower redshift clusters is already in place in Cl J0227-0421. Using the relationship of Andreon (2012), the
halo mass within $r_{500}$ corresponding to the composite corrected stellar mass calculated from the members of Cl J0227-0421 is
$M_{\Sigma\mathcal{M}_{s},\, 500}= 1.35\pm0.53 \times10^{14}$ $h_{70}^{-1}$ $M_{\odot}$ at $z\sim3.3$. However, to compare this
value fairly to the previous estimate it is necessary to correct the halo mass to that at the virial radius. This correction was done by modeling the halo mass profile as an NFW
profile with a concentration at the virial radius of $c_{vir}=2.5$ (Duffy et al.\ 2008). A correction factor of $c_{NFW}=1.39\pm0.48$ was determined by
the ratio of the total mass contained within $r_{vir}$ to that contained within $r_{500}$ as determined by the velocity dispersion. This correction factor was used to 
derive the final $z=3.3$ halo mass estimate coming from this method which is given in Table \ref{tab:massprop}. 

The halo mass limit placed on Cl J0227-0421 from the X-ray imaging data was performed as follows. To measure the X-ray flux limit at the position of Cl J0227-0421, all bright 
X-ray sources in the vicinity of the protocluster were optimally masked using the methodology presented
in Clerc et al.\ (2012). A bright source, identified as a local galaxy ($z=0.053$), lies within close proximity (1\arcmin) of the protocluster center, further compounding the 
difficulty of the measurement because the shot noise from this source is the overwhelming source of background noise in some parts of the adopted apertures. After masking, we integrated the 
count rate in concentric annuli and subsequently corrected for vignetting with the uncertainties derived through Poisson statistics. The measurement yields the total 
(summing three EPIC detectors: MOS1, MOS2, PN) count rate into the [0.5-2] keV band within physical aperture corresponding to 0.5 Mpc at $z\sim3.3$ and centered on the protocluster optical 
(i.e., number-weighted) position. These corrected count rates were then converted into a flux limit in the [0.5-2] keV band using a constant conversion factor of 
9$\times$10$^{-13}$ ergs cm$^{-2}$ counts$^{-1}$ (Adami et al.\ 2011), resulting in the limit quoted in \S\ref{halomass}. All random errors quoted for this method are 
Poissonian. This flux limit was converted into a rest-frame luminosity using the method described in \S\ref{halomass}, which resulted in a luminosity limit 
of $L_{X,[0.1-2.4 \, \, \rm{keV}], rest} < 3.98\pm0.96\times10^{45}$ ergs s$^{-1}$. The relationship between the X-ray luminosity in the rest-frame [0.1-2.4] keV, 
as measured at $r_{500}$\footnote{The flux limit is actually measured at roughly $r_{200}$, but the correction between the luminosity at this radius and the luminosity at $r_{500}$ 
is negligible (see Piffaretti et al.\ 2011).} and the hydrostatic equilibrium mass is given as (Arnaud et al.\ 2010, Piffaretti et al.\ 2011)

\begin{equation}
h(z)^{-7/3}\left( \frac{L_{X,\, 500,\, \,[0.1-2.4] \, \rm{keV}, \,rest}}{10^{44}} \right) = c\left(\frac{\mathcal{M}_{X,\, 500}}{3\times10^{14} M_{\odot}}\right)^{\alpha}
\label{eqn:LXmass}
\end{equation}

\noindent where $h(z)=H(z)/H_{0}$, $\log(c)=0.274\pm0.037$, and $\alpha=1.64\pm0.12$. Just as the member dynamics are unlikely to be governed by the virial theorem, the 
protocluster ICM, if it exists, is almost certainly not going to be in hydrostatic equilibrium, and thus this method only serves to crudely place limits on the halo 
mass. Using the above formula and correcting to the virial radius with an NFW as with the previous method results in the hydrostatic halo mass limit 
given in Table \ref{tab:massprop}.

The galaxy overdensity, $\delta_{gal}$, calculated in \S\ref{halomass} within the ``effective'' radius of Cl J0227-0421 was transformed into the matter overdensity, 
$\delta_m$, via (Steidel et al.\ 1998):

\begin{equation}
1+b\delta_{m}=C(1+\delta_{gal})
\label{eqn:dm}
\end{equation}

\noindent and

\begin{equation}
C=1+f-f(1+\delta_{m})^{1/3}
\label{eqn:corrredshiftspace}
\end{equation}

\noindent where C is defined as factor to correct the observed volume for redshift space distortions such that $C=V_{apparent}/V_{true}$, where $V_{apparent}$ is the measured
volume and $V_{true}$ is the volume after correction. This factor, discussed extensively in Steidel et al.\ (1998), has a complicated dependence on both the magnitude and 
the directionality of the velocities of the member galaxies. As in Cucciati et al.\ (2014), we made the assumption here that the structure is under collapse, such that $C<1$ (i.e.,
the galaxies are compressed in redshift space) and that the velocities are isotropic. Under these assumptions, $f$ in Equation \ref{eqn:corrredshiftspace} can be approximated
as $\Omega_{m}^{4/7}(z)$, the matter density relative to critical of the universe at redshift $z$ (for further details see Lahav et al.\ 1991; Padmanabhan 1993; Steidel et
al.\ 1998). Solving the above system of equations results in a correction factor of $C=0.53^{+0.16}_{-0.09}$ and a matter overdensity of $\delta_{m}=2.79^{+1.62}_{-1.46}$. 
The latter value can be translated into a $z=0$ halo mass via (Chiang et al.\ 2013)

\begin{equation}
\mathcal{M}_{\delta_{m}, \, tot, \, z=0} = C_{e}(1+\delta_{m,e})\Omega_{m,0} \, \rho_{crit,0}V_{e}
\label{eqn:deltamass}
\end{equation}

\noindent where $V_{e}$ is the effective volume, i.e., (2$R_{e}$)$^{3}$, where $R_{e}$ is measured in comoving Mpc, and $C_{e}$ is an additional correction factor to account
for mass outside of the effective radius. In Chiang et al.\ (2013), $C_{e}$ was found to be 2.5; i.e., 40\% of the mass was contained within a box defined by $R_{e}$, and that
is the value we adopt here. In principle, under the assumptions made here it is necessary to decrease the observed $R_{e}$ along the line-

of-sight dimension (i.e., a smaller 
redshift window) by a factor $C$ to match the simulated volume that is absent of distortions due to peculiar velocities. However, the value of $\delta_{gal}$ for Cl J0227-0421
is essentially invariant with respect to the redshift window chosen (for $\Delta z_{spec} < 0.08$) and, given the large uncertainty in the value of $C$, both in the formal error and the number 
of assumptions, we make no attempt to correct our observed redshift window for this effect. An additional uncertainty in this calculation relates to the scatter of 
$\delta_{m}$ and $\mathcal{M}_{\delta_{m}, \, tot, \, z=0}$ of simulated clusters. Because of the relatively small number of $z=0$ clusters in the analysis of Chiang et al.\ (2013) at or 
exceeding the total mass of the predicted descendant of Cl J0227-0421, this scatter is highly uncertain. Because of this uncertainty, and because the inclusion of this scatter in the formal 
error on the total mass calculated from this method does not change the interpretation of our results, we chose to ignore it. The total halo mass derived in Equation \ref{eqn:deltamass} is then 
converted to a halo mass at the virial radius using the method defined in the previous mass calculations, resulting in the value given in Table \ref{tab:massprop}. 

\end{document}